\documentclass[12pt]{article}

\newfont{\twelvemsb}{msbm10 scaled\magstep1}
\newfont{\eightmsb}{msbm8}
\newfam\msbfam
\textfont\msbfam=\twelvemsb
\scriptfont\msbfam=\eightmsb
\catcode`\@=11
\def\Bbb{\ifmmode\let\next\Bbb@\else
  \def\next{\errmessage{Use \string\Bbb\space only in math mode}}\fi\next}
\def\Bbb@#1{{\fam\msbfam{{#1}}}}

\newcommand{\be}{\begin{equation}}
\newcommand{\ee}{\end{equation}}
\newcommand{\ba}{\begin{eqnarray}}
\newcommand{\ea}{\end{eqnarray}}

\newcommand{\NP}[1]{Nucl.\ Phys.\ {\bf #1}}
\newcommand{\PL}[1]{Phys.\ Lett.\ {\bf #1}}

\newcommand{\CMP}[1]{Comm.\ Math.\ Phys.\ {\bf #1}}

\newcommand{\MPL}[1]{Mod.\ Phys.\ Lett.\ {\bf #1}}
\newcommand{\IJMP}[1]{Int.\ J.\ Mod.\ Phys.\ {\bf #1}}

\begin{document}
\sloppy
\renewcommand{\thefootnote}{\fnsymbol{footnote}}

\newpage
\setcounter{page}{1}

\vspace{0.7cm}
\begin{flushright}
DCPT-02/69\\
EMPG-02-20\\
May 2003
\end{flushright}
\vspace*{1cm}
\begin{center}
{\bf  Exact conserved quantities on the cylinder I: conformal case.}\\
\vspace{1.8cm}
{\large D.\ Fioravanti $^a$ and M.\ Rossi $^b$ \footnote{E-mail:
Davide.Fioravanti@durham.ac.uk, M.Rossi@ma.hw.ac.uk}}\\ \vspace{.5cm}
$^a${\em Department of Mathematical Sciences,
South Road, Durham DH1 3LE,
England} \\

\vspace{.3cm}
$^b${\em Department of Mathematics, Heriot-Watt University, Edinburgh EH14 4AS,
Scotland} \\
\end{center}
\vspace{1cm}

\renewcommand{\thefootnote}{\arabic{footnote}}
\setcounter{footnote}{0}

\begin{abstract}
{\noindent The} nonlinear integral equations describing the spectra of the left and right (continuous) quantum KdV equations on the cylinder are derived from integrable lattice field theories, which turn out to allow the Bethe Ansatz equations of a twisted ``spin \hspace{0.1cm} $-1/2$'' chain. A very useful mapping to the more common nonlinear integral equation of the twisted continuous spin $+1/2$ chain is found. The diagonalization of the transfer matrix is performed. The vacua sector is analysed in detail detecting the primary states of the minimal conformal models and giving integral expressions for the eigenvalues of the transfer matrix. Contact with the seminal papers \cite{BLZ, BLZ2} by Bazhanov, Lukyanov and Zamolodchikov is realised. General expressions for the eigenvalues of the infinite-dimensional abelian algebra of local integrals of motion are given and explicitly calculated at the free fermion point.

\end{abstract}

\vspace{1cm}


\newpage

\section {Introduction}

An integrable system in $1+1$ dimensions is characterized by the existence of a number of integrals of motion in involution equal to the number of degrees of freedom. Consequently, the exact computation (of the eigenvalues) of these commuting observables is a very important problem. In this respect, an important and widely used method goes through (an integrable discretisation of the model and) the diagonalization of the discrete integrals of motion by a Bethe Ansatz technique. At the end one is left with the problem of finding the continuous limit of the eigenvalues of the (discretised) theory. In order to achieve this goal, a powerful analytic method was introduced by Destri and de Vega for the periodic sine-Gordon theory on a cylinder \cite{DDV}. Although we will follow the way depicted in \cite{DDV}, we want to mention that similar results were first obtained by \cite {KP} in the case of the critical behaviour of the 6- and 19-vertex models by using a different technique. We do prefer to use the procedure of Destri and de Vega since it has been successfully applied to the calculation of ground state energy \cite {DDV1} and first excited states energies \cite{FMQR} of sine-Gordon model, being our aim in the second paper the treatment of massive perturbation of Conformal Field Theories (cfr. \cite{FRa} as review paper on the subject).

In the paper \cite {FR} we have formulated a generalisation of the Algebraic Bethe Ansatz to diagonalise the transfer matrices of two lattice models, obtained after discretisation (on a space lattice) of two - left and right - quantum (m)KdV equations. The lattice formulation of one lattice (m)KdV was already performed by Kundu \cite{KUN}, stressing the necessity of modifying the usual Yang-Baxter relation to take into account the non-ultralocality of the theory. The integrability structure was found in a braided version of the Yang-Baxter relation, which has been also attracting mathematical interest \cite{HLK} in itself. Afterwards, we have found the Bethe equations, which may be thought of as those of a ``spin \hspace{0.2cm} $-1/2$'' (``wrong'' sign) XXZ chain with a twist. The origin of the twist in the framework of \cite {FR} is in some sense {\it dynamical}, i.e. automatically generated by the theory and depending on the number of Bethe roots, almost like in \cite{FT} and unlike in \cite {FMQR}. It is also worth mentioning that a space {\it and} time discretisation of the KdV was proposed in \cite{V} and analysed by using the Baxter $TQ-$system approach \cite{BAX}: these discrete-discrete integrable systems are indeed important also for their continuous parents \cite{FV}.

In this article we address the problem of writing nonlinear integral equations governing the spectrum of the left and right (continuous) quantum (m)KdV theories. In other words, the eigenvalues of the transfer matrix - which encodes all the local and nonlocal conserved charges - are expressed as nonlinear functionals of the solutions of the aforementioned nonlinear integral equations. Besides the importance in Statistical Field Theory, they can be also interpreted as explicit expressions for spectral determinants of ${\cal P}{\cal T}$- symmetric spectral problems \cite {BBM,DT}. This procedure will eventually give the eigenvalues of the local integrals of motion of the quantum KdV theory. After \cite {IS}, it became well known that these local integrals of motion represent one series of local commuting charges of conformal field theories without extended symmetries \cite {FFr,FRS} (cfr. \cite {FR,BLZ,FS} for details concerning the matrix Lax formulation). In addition, this series is exactly that which is still conserved, after suitable modification, if the conformal field theory is perturbed by its $\Phi_{(1,3)}$ operator. Actually, applying the investigations of the present paper and of \cite {FR} to the setup elaborated in \cite {FRS} about the $A_2^{(2)}$ KdV theory, it is only a matter of calculations to extend the subsequent results to the only other series \cite {FFr}, characterising the only other integrable perturbations. The very intriguing feature of the conformal formulation in terms of integrable hierarchies is that the conformal monodromy matrix already contains the perturbing field and the screening operator \cite{FRS}. A similar discretisation should be also allowed in conformal field theories with extended symmetry algebra (e.g. for ${\cal W}$ algebras it is of basic importance the setting of \cite{FL}; cfr. \cite{BHK} for development in a peculiar case).

We remark that similar results have been obtained in pioneering works \cite {BLZ,BLZ2} by Bazhanov, Lukyanov and Zamolodchikov from a different starting point and via a different approach. They define directly in the continuous field theory the transfer matrix as an operator series acting on Virasoro modules (for $0 < \beta^2 < 1/2$). On the other hand, we intend to create a bridge between conformal field theories and the powerful algebraic Bethe Ansatz formulation of lattice KdV \cite{FR, KUN} -- based on this interesting generalisation of the Yang-Baxter algebra --, showing that the continuum field theory limit can be recovered in the braided case as well. As particular consequence, in the second paper we will be able to formulate the $\Phi_{(1,3)}$ perturbation of conformal field theory within the same framework of \cite{FR} by merging two KdV theories. In any case, the monodromy matrix of \cite{BLZ,BLZ2} is different from the na\"{\i}ve continuous limit \cite {FR} of our monodromy matrices, which, unlike in \cite {BLZ,BLZ2}, are not solutions of the usual Yang-Baxter equation.

In section 2 we summarise the main results on (m)KdV theories obtained in \cite {FR}. In section 3, starting from the Bethe equations of \cite {FR}, we write the nonlinear integral equation for each vacuum of the left and right quantum KdV equations. In section 4 we use this equation to calculate up to quadratures the energy of the vacua. In section 5 we
apply the nonlinear integral equation to find exact expressions for the continuous limit of the vacuum eigenvalues of the transfer matrix. From the asymptotic expansion of these eigenvalues we obtain the local abelian charges, mentioned before as characterising the $\Phi_{(1,3)}$ perturbation. In addition, we have solved exactly the theory in two cases: at the free fermion point $\beta^2=\frac{1}{2}$ and in the limit of infinite twist ($\tilde{\omega}\rightarrow+\infty$). Some results are also compared with results and conjectures in \cite {BLZ,BLZ2}. In Section 6 we summarise our work describing also possible further applications.

\section{Quantum lattice (m)KdV theory}
\setcounter{equation}{0}

In a previous paper \cite{FR} we considered the left and right mKdV
equation, respectively
\be
\partial _{\tau}v=\frac {3}{2}v^2v^\prime +\frac {1}{4}v^{'''}
\quad , \quad   \partial _{\bar\tau}\bar v=\frac {3}{2}\bar v^2\bar v^\prime
+\frac {1}{4}\bar v^{'''} \, , \label {mkdv}
\ee
for the fields $v\equiv-\varphi^\prime$ and $\bar v \equiv-\bar \varphi ^\prime$, defined as
spatial derivatives of quasi-periodic Darboux fields $\varphi$, $\bar \varphi$ in
the interval $y$, $\bar y \in[\, 0\, ,\, R\, ]$. The quantizations of the
Darboux fields are the Feigin-Fuks bosons $\phi$, $\bar \phi$ \cite{FF}, which satisfy the
commutation relations   \begin{equation}
[\phi(y) \, , \, \phi  (y^\prime)]=-{\frac {i\pi \beta ^2}{2}}{\mbox s}\left (\frac{y-y^\prime}{R}\right)
\quad , \quad   [\bar \phi (\bar y) \, , \, \bar \phi
(\bar y^\prime )]={\frac  {i\pi \beta ^2}{2}}{\mbox s} \left (\frac {\bar y-\bar
y^\prime}{R}\right) \, ,
\label{phi}
\ee
where $0<\beta ^2<1$ and ${\mbox s}(z)$ is the quasi-periodic extension of the sign
function:
\[
{\mbox s}(z)=2n+1 \quad , \quad n<z<n+1 \quad ;  \quad {\mbox s}(n)=2n \quad , \quad n
\in {\Bbb Z} \, .
\]
In terms of a discretisation of the Feigin-Fuks
bosons, $\phi_m\equiv \phi\left (m\frac{R}{2N}\right)$, $\bar \phi_m\equiv
\bar \phi \left (m\frac {R}{2N}\right )$, we defined the $N$-periodic operators
$V_m^{\pm}$ living on a $N$-site lattice with spacing $\Delta $ and length
$R=N\Delta$:
\ba
V^-_m\equiv \frac {1}{2}\left [ (\phi _{2m-1}-\phi_{2m+1})+(\phi _{2m-2}-\phi  _{2m})
-(\bar \phi _{2m-1}-\bar \phi_{2m+1})+(\bar  \phi _{2m-2}-\bar \phi _{2m})\right]
\label{Ccorr1} \\
V^+_m\equiv \frac {1}{2}\left [(\bar \phi _{2m-1}-\bar \phi _{2m+1})+ (\bar
\phi_{2m-2}-\bar \phi_{2m})-(\phi _{2m-1}-\phi  _{2m+1})+(\phi_{2m-2}-\phi _{2m}) \right] .
\label{Ccorr2}
\ea
These operators are the quantum counterparts of the discretisation of the
mKdV variables $v$ and $\bar v$ (\ref {mkdv}) and satisfy the non-ultralocal
commutation relations, first introduced in \cite {KUN},
\begin{eqnarray} &&[V^+_m\, , \,
V^+_n]=\frac {i\pi \beta ^2}{2}(\delta _{m-1,n}-\delta _{m,n-1})
\, , \nonumber \\ &&[V^-_m\, , \, V^-_n]=-\frac {i\pi \beta ^2}{2}(\delta
_{m-1,n}-\delta _{m,n-1}) \, , \label{vrel2} \\ &&[V^+_m\, , \,
V^-_n]=-\frac {i\pi \beta ^2}{2}(\delta _{m-1,n}-2\delta
_{m,n}+\delta _{m,n-1}) \,  \nonumber \, . \end{eqnarray}

In terms of $V_m^\pm$ we now define (adopting a different normalisation with respect to \cite {FR}) the {\it left} and {\it right} conformal
monodromy matrices
\ba  M(\alpha)&=&e^{\frac {i\pi \beta ^2}{2}}L_{N}(\alpha )\ldots L_{1}(\alpha ) \,
, \label {leftmon} \\  \bar M(\alpha)&=&e^{-\frac {i\pi \beta ^2}{2}}\bar L_{N}(\alpha)\ldots \bar
L_{1}(\alpha) \, ,
\label {rightmon}
\ea
where left Lax operators $L_m(\alpha)$ and right Lax operators
$\bar L_m(\alpha)$ are given by
\begin{equation}
L_{m}(\alpha) \equiv e^{-\frac {i\pi \beta ^2}{4}}\left (
\begin{array}{cc} e^{-iV_m^-} & \Delta e^{\alpha}  e^{iV_m^+} \\ \Delta e^{\alpha}
e^{-iV_m^+}&  e^{iV_m^-}\\ \end{array} \right )   ,  \bar L_{m}(\alpha)
\equiv e^{\frac {i\pi \beta ^2}{4}}\left ( \begin{array}{cc} e^{-iV_m^+} &  \Delta e^{-\alpha}   e^{iV_m^-}\\
\Delta e^{-\alpha}  e^{-iV_m^-} &   e^{iV_m^+}\\  \end{array} \right ) \, .
\label{lax}
\end{equation}

As a consequence of non-ultralocal relations (\ref {vrel2}),
monodromy matrices (\ref {leftmon}, \ref {rightmon}) satisfy braided Yang-Baxter relations (see, for instance, relation (4.4) of \cite {FR}), first introduced in \cite {KUN} and \cite {HLK}. The commuting transfer matrices associated to them generate, restoring the continuum, one series of local commuting charges of, respectively, left and right conformal field theories.

Despite the complication coming from the presence of a braided Yang-Baxter relation, we have derived in \cite {FR} the Bethe equations for the monodromy matrices (\ref {leftmon}, \ref {rightmon}). They read:
\begin{equation}
\prod _{{\stackrel{r=1}{r\not= s}}}^l \frac {\sinh (\alpha _s -\alpha _r +i\pi \beta ^2)}{\sinh (\alpha _s -\alpha _r -i\pi \beta ^2)}=\left [\frac
{\sinh \left (\alpha _s -\frac {i\pi \beta ^2}{2}-\varepsilon \ln \frac {R}{N}\right)}
{\sinh \left (\alpha _s +\frac {i\pi \beta ^2}{2}-\varepsilon\ln \frac {R}{N}\right)}
\right]^{N/2} e^{\varepsilon i\pi \beta ^2 \left (\frac {N}{2}+2l\right )}\, .
\label{Bethe}
\end{equation}
The equations may be formally interpreted as being those of a twisted ``spin \hspace{0.3cm} $-1/2$'' XXZ chain, for a sign is the only difference with respect to the usual twisted spin $+1/2$ chain equations. Indeed, $l$ is the number of the Bethe roots and the parameter $\varepsilon $ is equal to $-1$ for the left conformal case (monodromy matrix (\ref {leftmon})) and to $+1$ for the right conformal case (monodromy matrix (\ref {rightmon})). The connection with notations used in formul{\ae} (6.35, 6.47) of \cite {FR} is given by $q=e^{-i\pi \beta ^2}$ and $\lambda _r=e^{\alpha _r}$.

In the same paper \cite {FR} we wrote also the eigenvalues of the left and right lattice transfer matrices on the Bethe states. Using the same notations as in (\ref {Bethe}) such eigenvalues read
\be
{\Lambda}_N(\alpha )={\Lambda}^+_N(\alpha )+{\Lambda}^-_N(\alpha ) \, ,
\label {tra}
\ee
where
\be
{\Lambda}^{\pm}_N(\alpha )=
e^{\mp \varepsilon i \pi \beta ^2 l} \prod\limits _{r=1}^{l} \frac {\sinh (\alpha -\alpha _r\pm i\pi \beta ^2)}{\sinh (\alpha -\alpha _r) } \rho _N^{\pm}(\alpha)
\label {tra1}
\ee
and
\be
\rho _N ^{\pm}(\alpha )=
e^{\mp \varepsilon \frac {i\pi \beta ^2N}{4}} \left (\frac {2R}{N}e^{-\varepsilon \alpha}\right)^{N/2}\sinh \left ( \varepsilon \alpha \pm \varepsilon \frac {i\pi \beta ^2}{2} -\ln \frac {R}{N}\right ) ^{N/2} \, .
\label {rho}
\ee

In this paper we will perform the continuous limit of these quantum lattice (m)KdV theories, with the aim of studying quantum (m)KdV theories.

\section {The nonlinear integral equation for the vacua sector}
\setcounter{equation}{0}

We want to write a nonlinear integral equation equivalent to Bethe equations (\ref {Bethe}) for the particular solution minimising the energy (the vacuum). We introduce the analytic function in the strip $|{\mbox {Im}}\, x |<\, {\mbox {min}}\, \{\zeta, \pi -\zeta\} \,  , \, \, 0\leq \zeta < \pi $ ($\ln$ is taken in the fundamental branch)
\begin{equation}
\phi (x , \zeta)\equiv i\ln \frac {{\mbox {sinh}} (i\zeta+x)}{{\mbox {sinh}} (i\zeta-x)}\, , \label {funz}
\end{equation}
to define, when $N\in 4{\Bbb N}$, the counting function,
\begin{equation}
Z_N(x)\equiv \frac {N}{2}\phi \left (x -\varepsilon \ln \frac {R}{N}, \frac {\pi }{2}\beta ^2 \right )+\sum _{r=1}^{l}\phi (x -\alpha _r, \pi \beta ^2)+2\pi \beta ^2 \varepsilon \left (l+\frac {N}{4}\right )  \, , \label{count0}
\end{equation}
in terms of which the Bethe equations (\ref {Bethe}) have the logarithmic form
\begin{equation}
Z_N(\alpha _s)=\pi (2I_s-l-1 ) \, , \quad   I_s\in {\Bbb Z} \, . \label {cond1}
\end{equation}
First of all, we are interested in the vacuum solutions of the Bethe equations: all the $\alpha _s$ are real and $Z_N(\alpha _s)$ are equal to all the numbers of the form (\ref {cond1}) between $Z_N(-\infty)$ and $Z_N(+\infty)$, i.e. there are no {\it holes} (cfr. sub-section 5.2 for further details). It follows from the definition (\ref {count0}) that these two extremes are given by
\be
Z_N(\pm\infty)=\pm\frac {N}{2}(\pi-\pi \beta ^2)\pm l(\pi-2\pi \beta ^2)+2\pi \beta ^2 \varepsilon \left (l+\frac {N}{4}\right )  \, . \label {extr}
\ee
This solution minimises, for the range of values of $\beta^2$
\be
\frac {1}{3}<\beta ^2 <\frac {2}{3} \label {range} \, ,
\ee
the left ($\varepsilon =-1$) and right ($\varepsilon =1$) hamiltonian eigenvalues, defined as {\it logarithmic derivatives} of the transfer matrix eigenvalues, respectively
\ba
H_{\varepsilon}(\{\alpha _r\},R,N)&=&-\frac {\Delta \varepsilon}{2\sin \pi \beta ^2}\left [e^{-2\varepsilon\alpha}\frac {\partial}{\partial \alpha}G^-_N(\alpha)\left |_{\alpha=\varepsilon \ln \frac {R}{N}-\frac {i\pi \beta ^2}{2}}\right.+ \right. \nonumber \\
&+& \left. e^{-2\varepsilon\alpha}\frac {\partial}{\partial \alpha}G^+_N(\alpha)\left |_ {\alpha=\varepsilon \ln \frac {R}{N}+\frac {i\pi \beta ^2}{2}}\right. \right]
\, , \label{ham}
\ea
where
\be
G^\pm _N(\alpha)=\ln [ \rho _N ^{\pm}(\alpha)^{-1} {\Lambda}_N(\alpha) ] \, .
\label{Gpm}
\ee
On the other hand, the vacuum solution is known to minimise in the region (\ref {range}) the lattice sine-Gordon Hamiltonian eigenvalues $H_{LSG}(\{\alpha _r\},R,N)$ \cite {BOG}; left and right hamiltonians eigenvalues (\ref {ham}) can be regarded as the conformal limit of
$H$:
\be
H_{\varepsilon}(\{\alpha _r\},R,N)=\lim _{\Theta \rightarrow +\infty}H_{LSG}(\{\alpha _r-\varepsilon \Theta -\varepsilon \ln R/N \},R,N) \, .
\ee
Supposing $\beta ^2$ in the region (\ref {range}),
we now want to write a nonlinear integral equation for the vacuum counting function in the continuous limit:
\be
N\rightarrow +\infty \, , \quad R \, \, {\mbox {fixed}} \, . \label{contlim}
\ee
Besides, this limit means, because of (\ref {extr}), that also $l\rightarrow +\infty$ and therefore, even if we put
\be
\beta ^2 =\frac {p}{p^\prime} \, , \label{beta}
\ee
with $p<p^\prime$ coprimes, it turns out to be ill-defined on the exponential factor of (\ref {Bethe}). Therefore, we slightly modify the method in \cite{FT} and we need to consider different ways for $l$ to go to $\infty$ according to the sub-sequences
\be
l=np^\prime+\kappa \quad , \quad 0\leq \kappa \leq p^\prime -1 \, ,
\label {par1}
\ee
where $n \rightarrow +\infty$ and $\kappa$ is kept fixed.
The utility of this limit procedure is that we can redefine the counting function (\ref {count0}) up to suitable multiples of $2\pi$ to obtain
\be
Z_N(x)=\frac {N}{2}\phi \left (x -\varepsilon \ln \frac {R}{N}, \frac {\pi}{2}\beta ^2 \right )+ \sum _{r=1}^{l}\phi \left (x -\alpha _r, \pi \beta ^2 \right )+2\pi \frac {N}{4}\varepsilon \left (\beta ^2-1\right)+2\pi \varepsilon \omega  \, , \label{count1}
\ee
where $\omega$ is finite in the continuous limit (the double braces denote the fractional part):
\be
\omega =\left \{ \left \{ \frac {p\kappa}{p^\prime} \right \}\right \} \, . \label{omega}
\ee
The presence of the constant divergent term in definition (\ref {count1}) will be made clear in the following.

Now, we can use standard techniques (see \cite {DDV}, \cite {DDV1}, \cite {FMQR}) to rewrite a sum over real Bethe roots of a function $f_N$ with no poles on the real axis as follows \footnote {We are fixing $l$ to be even in the following (i.e. $e^{iZ_N(\alpha _s)}=-1$), as the case $l$ odd can be treated along the same lines and in the end does not add new information.}:
\be
\sum _{r=1}^{l}f_N(\alpha _r)=-\int _{-\infty}^{+\infty}\frac {dx}{2\pi} f^\prime _N (x)Z_N(x)+2\int _{-\infty}^{+\infty}\frac {dx}{2\pi}f^\prime _N(x){\mbox {Im}}\ln \left [1+e^{iZ_N(x+i0)}\right ] \, . \label {sumf1}
\ee

The use of (\ref {sumf1}) in (\ref {count1}) gives
\begin{eqnarray}
&&Z_N(x)=\frac {N}{2}\phi \left(x -\varepsilon \ln \frac {R}{N}, \frac {\pi}{2}\beta ^2\right) +2\pi \frac {N}{4}\varepsilon \left (\beta ^2-1\right)+2\pi \varepsilon \omega +
\label{count2} \\
&+&\int _{-\infty}^{+\infty}\frac {dy}{2\pi} \phi ^\prime (x-y, \pi \beta ^2)Z_N(y)-2\int _{-\infty}^{+\infty}\frac {dy}{2\pi}\phi ^\prime (x - y,\pi \beta ^2 ){\mbox {Im}}\ln \left [1+e^{iZ_N(y+i0)}\right ] \nonumber
\, .
\end{eqnarray}

After defining:
\be
K(x)\equiv\frac {1}{2\pi}\phi ^{\prime}\left (x, \pi \beta ^2 \right )\, , \quad \Phi (x)\equiv   \phi \left(x, \frac {\pi }{2}\beta ^2\right) \, ,  \quad
L_N(x)\equiv{\mbox {Im}}\ln \left [1+e^{iZ_N(x+i0)}\right ]\, , \label {not}
\ee
the Fourier transform of relation (\ref {count2}) is
\be
\hat Z_N(k)=\frac {N}{2}\frac {\hat \Phi (k)e^{-i\varepsilon k\ln \frac {R}{N}}}{1-\hat K(k)}-2\frac {\hat K(k)}{1-\hat K(k)}
\hat L_N(k)+4\pi ^2 \varepsilon \left [ \omega  + \frac {N}{4}\left (\beta ^2-1\right )\right ]
\frac  {\delta (k)}{1-\hat K(k)} \, , \label {count3}
\ee
where we have used the notation:
\be
\hat f(k)\equiv \int _{-\infty}^{+\infty} dx e^{-ikx} f(x) \, .
\label {fourier}
\ee

Explicitly, the Fourier transforms of the first two functions in (\ref {not}) are:
\be
\hat \Phi (k)= \frac {2\pi}{ik}\frac {\sinh k\left (\frac {\pi}{2}-\frac {\pi}{2}\beta ^2\right)}{\sinh \frac {\pi}{2}k} \quad , \quad \hat K(k)=\frac {\sinh k\left (\frac {\pi}{2}-\pi \beta ^2 \right)}{\sinh \frac {\pi}{2}k} \, . \label {four1}
\ee
With the definitions
\be
\gamma\equiv \pi (1-\beta ^2) \, , \quad \hat G(k) \equiv -\frac {\hat K(k)}{1-\hat K(k)}=\frac {\sinh k\left (\frac {\pi}{2}-\gamma \right)}{2 \cosh k \frac {\gamma}{2} \sinh k\left (\frac {\pi}{2}-\frac {\gamma}{2}\right)} \, , \quad \tilde \omega =\frac {\pi }{\beta ^2}\omega  \, , \label {G}
\ee
equation (\ref {count3}) can be written in the following way:
\be
\hat Z_N(k)=\frac {N}{2}\frac {\hat \Phi (k)e^{-i\varepsilon k\ln \frac {R}{N}}}{1-\hat K(k)}+2\hat G(k)\hat L_N(k)+2\pi  \varepsilon \left [ \tilde \omega  + \frac {N}{4}\pi \left (1-\frac {1}{\beta ^2} \right )\right ]
{\delta (k)} \, . \label {count4}
\ee

We now want to Fourier antitransform and to perform the continuous limit.
Therefore, it is necessary to work out in such a limit the behaviour of
the $N$-dependent term
\ba
F_N(x)&\equiv& \int _{-\infty}^{+\infty} \frac {dk}{2\pi} e^{ikx}
\left (\frac {N}{2}\frac {\hat \Phi (k)e^{-i\varepsilon k\ln \frac {R}{N}}}{1-\hat K(k)}\right ) = \nonumber \\
&=&\frac {\pi N}{2i}\int _{-\infty}^{+\infty} \frac {dk}{2\pi k}
e^{ik\left (x-\varepsilon \ln \frac {R}{N}\right)} \frac {\sinh k\left (\frac {\pi}{2}-\frac {\pi }{2}\beta ^2\right)}{\sinh k \frac {\pi }{2}\beta ^2 \cosh k\left (\frac {\pi}{2}-\frac {\pi }{2}\beta ^2\right)} \, . \label {forc}
\ea
This expression obviously diverges in the continuous limit. To define a finite counting function in such a limit we perform a $N$-dependent shift $\Lambda _N$ in $x$ and define the conformal counting function $Z(x)$ as
\be
Z(x)\equiv \lim _{\stackrel {N\rightarrow +\infty}{N\Delta =R}} Z_N(x+\Lambda _N) \, . \label {confcount}
\ee
We have to choose $\Lambda _N$ in such a way that
\be
F_N(x+\Lambda _N)=\frac {\pi N}{2i}\int _{-\infty}^{+\infty} \frac {dk}{2\pi k}
e^{ik\left (x-\varepsilon \ln \frac {R}{N}+\Lambda _N\right)} \frac {\sinh k\left (\frac {\pi}{2}-\frac {\pi}{2}\beta ^2\right)}{\sinh k \frac {\pi}{2}\beta ^2 \cosh k\left (\frac {\pi}{2}-\frac {\pi}{2}\beta ^2\right)} \,  \label {forc2}
\ee
is finite in the continuous limit. A convenient way to evaluate (\ref {forc2}) is the residues method. The poles of the integrand lie at $k=0$ and on the imaginary axis. Since we are in the region (\ref {range}),
the poles on the imaginary axis with smaller modulus are at $k=\pm i\pi/\gamma$. We choose $\Lambda _N $ as follows
\be
-\varepsilon \ln \frac {R}{N}+\Lambda _N = \chi \frac {\gamma}{\pi} \ln \left [\frac {A}{N} \sin \frac {\pi \beta ^2}{2(1-\beta ^2)}\right ] \, ,
\ee
where $\chi$ can be equal to $\pm 1$ and $A$ is an arbitrary
positive constant which reflects the translation invariance in
$x$ of conformal field theories.

If $\chi =-1$, then in the continuous limit $-\varepsilon \ln \frac {R}{N}+\Lambda _N \rightarrow +\infty$: hence, to evaluate the integral for fixed $x$ we can choose a contour closing in the upper $k$-complex half-plane and containing a small semicircle surrounding the pole at $k=0$. Since the contribution from the imaginary poles are damped by the exponential factor containing $\Lambda _N$, the leading term is given by the residue at the pole with smallest modulus, i.e. $k=i\pi/\gamma$, minus the contribution of the semicircle around $k=0$:
\be
N\rightarrow +\infty \, , \quad F_N(x+\Lambda _N)=-Ne^{-\frac {\pi}{\gamma}x} e^{\ln \frac {A}{N}}+\frac {N}{4}\pi \left (\frac {1}{\beta ^2}-1\right ) +o(N^0)
\, , \label {forc3}
\ee
where $o(z)$ means ``order less than $z$''.

If $\chi =1$, then in the continuous limit $-\varepsilon \ln \frac {R}{N}+\Lambda _N\rightarrow -\infty$ and we consider a contour closing in the lower $k$-complex half-plane and containing a small semicircle surrounding the pole at $k=0$. As before, we take into account only the contribution from the residue at the pole with smallest modulus, $k=-i\pi/\gamma$, and we subtract the contribution from the semicircle around $k=0$:
\be
N\rightarrow +\infty \, , \quad F_N(x+\Lambda _N)=Ne^{\frac {\pi}{\gamma}x} e^{\ln \frac {A}{N}}-\frac {N}{4}\pi \left (\frac {1}{\beta ^2}-1\right ) +o(N^0)\, .
\label {forc3'}
\ee
Collecting these formulas we get
\be
N\rightarrow +\infty \, , \quad F_N(x+\Lambda _N)=
A \chi e^{\chi \frac {\pi}{\gamma}x}-\chi \frac {N}{4}\pi \left (\frac {1}{\beta ^2}-1\right ) +o(N^0) \, .
\label {forc4}
\ee
We now make the choice
\be
\chi =- \varepsilon \, , \label{choice}
\ee
and insert (\ref {forc4}) in the continuous limit of the Fourier antitransform of (\ref {count4}) shifted by $\Lambda _N$. The constant divergent term disappears and we obtain for the conformal counting function (\ref {confcount}) the following equation:
\be
Z(x)=- A\varepsilon e^{- \varepsilon \frac {\pi}{\gamma}x}+
2\int_{-\infty}^{+\infty}{dy}\, G(x-y){\mbox {Im}}\ln \left [1+e^{iZ(y+i0)}\right ]+\varepsilon \tilde \omega \, ,
\label{ddv}
\ee
where $G(x)$ is the Fourier antitrasform of $\hat G(k)$ (\ref {G})
\be
G(x)=\int _{-\infty}^{+\infty} \frac {dk}{2\pi}e^{ikx}\frac {\sinh k\left (\frac {\pi}{2}-\gamma \right)}{2 \cosh k \frac {\gamma}{2} \sinh k\left (\frac {\pi}{2}-\frac {\gamma}{2}\right)} \, . \label{G(x)}
\ee

{\bf Remark 1} In writing (\ref {count3}) we have paid the necessary attention to the zero modes. Besides, we could more correctly derive equation (\ref {ddv}) working out our calculations on the derivative of $Z(x)$ \cite {DDV2}, but we think that this derivation is simpler, once all the subtleties are under control.

{\bf Remark 2} Although equation (\ref {ddv}) has been found for
$1/3<\beta ^2 < 2/3$, yet it can be considered without problems in the whole region $0<\beta ^2 <1$, defining in such way, by analytic continuation, a state in the continuous limit which minimises the energy.

{\bf Remark 3} In fact, equation (\ref {ddv}) coincides with the nonlinear integral equation for the spin $+1/2$ XXZ chain which shows Bethe equations (\ref {Bethe}) with the l.h.s. inversed and $\beta ^2=\frac {\gamma}{\pi}$.
Hence, the mapping $\beta^2 \rightarrow 1-\beta ^2$ relates the twisted spin $+1/2$ and spin $-1/2$ XXZ chains. This connection between spin $+1/2$ and spin $-1/2$ XXZ chains was already conjectured in \cite {FT}, but only in the case of infinite volume.

{\bf Remark 4} When $\varepsilon =-1$, we can make contact with Bazhanov, Lukyanov and Zamolodchikov's paper \cite {BLZ2}. Indeed, equation (\ref {ddv}) coincides with equation (3.8) of \cite {BLZ2}, derived for $0<\beta ^2 <1/2$, if we choose $A=2M\cos \frac {\pi \beta ^2}{2(1-\beta ^2)}$, where
\be
M=\frac {\Gamma \left ( \frac {\beta ^2}{2-2\beta ^2}\right )\Gamma \left ( \frac {1-2\beta ^2}{2-2\beta ^2}\right )}{{\sqrt {\pi}}}\Gamma (1-\beta ^2)^{\frac {1}{1-\beta ^2}} \, , \label {M}
\ee
and if we identify $p$ of \cite {BLZ2} with $\omega/2$. Such an $A$ is positive, as it has to be, in the whole region $0<\beta ^2 <1$. Therefore, in spite of the differences highlighted in the Introduction, (\ref {ddv}) with this choice of $A$ can be regarded as an extension of (3.8) of \cite {BLZ2} to the domain $0<\beta ^2 <1$.

\medskip

Equation (\ref {ddv}) is the nonlinear integral equation describing the $\tilde \omega$-vacuum of the continuous theory. Its counting function solution allows us to evaluate the infinite continuous limit of a sum over the Bethe roots of the $\tilde \omega$-vacuum resetting formula (\ref {sumf1}) as 
\be
\lim _{\stackrel {N\rightarrow +\infty}{N\Delta =R}} \sum _{r=1}^{l}f_N(\alpha _r)=-\int _{-\infty}^{+\infty}\frac {dx}{2\pi} f^{\prime} (x)Z(x)+2\int _{-\infty}^{+\infty}\frac {dx}{2\pi}f^{\prime}(x){\mbox {Im}}\ln \left [1+e^{iZ(x+i0)}\right ] \, , \label {sumf2}
\ee
where we have defined the continuous limit
\be
f(x)\equiv \lim _{\stackrel {N\rightarrow +\infty}{N\Delta =R}} f_N(x+\Lambda _N)
\, . \label{fcont}
\ee

Relation (\ref {sumf2}) contains two different contributions; we will call them, for the sake of brevity, the bulk term and the finite size correction term. They can be identified inserting in the first term of the r.h.s of (\ref {sumf2}) the expression for $Z(x)$ coming from (\ref {ddv}):
\ba
&&\lim _{\stackrel {N\rightarrow +\infty}{N\Delta =R}} \sum _{r=1}^{l}f_N(\alpha _r)=2\int _{-\infty}^{+\infty}\frac {dx}{2\pi}f^\prime (x){\mbox {Im}}\ln \left [1+e^{iZ(x+i0)}\right ] +  \label {sumf3} \\
&+&\int _{-\infty}^{+\infty}\frac {dx}{2\pi} f^\prime  (x)\left \{
A \varepsilon e^{- \varepsilon \frac {\pi}{\gamma}x }-2\int_{-\infty}^{+\infty}{dy}\, G(x-y){\mbox {Im}}\ln \left [1+e^{i Z(y+i0)}\right ] - \varepsilon \tilde \omega \right \}\, .  \nonumber
\ea
We can arrange this expression
\be
\lim _{\stackrel {N\rightarrow +\infty}{N\Delta =R}}\sum _{r=1}^{l}f_N(\alpha _r)=f_b+2\int _{-\infty}^{+\infty}\frac {dx}{2\pi} f^\prime (x)\int _{-\infty}^{+\infty}dy [\delta (x-y)-G(x-y)]{\mbox {Im}}\ln \left [1+e^{iZ(y+i0)}\right ] \, , \label {sumf4}
\ee
where $f_b$ indicates the bulk term
\be
f_b=\int _{-\infty}^{+\infty}\frac {dx}{2\pi} f^\prime (x)
\left [A \varepsilon e^{- \varepsilon \frac {\pi}{\gamma}x }
-\varepsilon \tilde \omega  \right ] \, , \label {bulk}
\ee
and the other term gives usually the finite size correction.
It is useful to write these terms using the Fourier transform of $f$, defined in (\ref {fourier}):
\be
\lim _{\stackrel {N\rightarrow +\infty}{N\Delta =R}} \sum _{r=1}^{l}f_N(\alpha _r)=f_b+\int _{-\infty}^{+\infty}\frac {dx}{\pi}
J_f(x) {\mbox {Im}}\ln \left [1+e^{iZ(x+i0)}\right ]  \, , \label {sumf5}
\ee
where
\be
J_f(x)=\int _{-\infty}^{+\infty} \frac {dk}{2\pi}e^{ikx} ik \hat J(k) \hat f(k)  \, , \quad \hat J(k)=\frac {\sinh k\frac {\pi}{2}}{2 \cosh k \frac {\gamma}{2} \sinh k\left (\frac {\pi}{2}-\frac {\gamma}{2}\right)} \, . \label {J}
\ee
In the next two sections we will evaluate the bulk term and the finite size correction term of the conformal hamiltonian and of the eigenvalues of the transfer matrix, both in the left (chiral) and in the right (antichiral) case.

\section{Finite size correction to the energy of the vacuum}

\setcounter{equation}{0}

The aim of this section is to extract, from the finite size correction to the vacuum energy, the conformal charge of the theory and the conformal weight, characteristic of the $\tilde \omega$-vacuum.

Let us start by writing the energy (\ref {ham}) of a Bethe state:
\[
H_{\varepsilon}(\{\alpha _r\},R,N)=\sum \limits _{r=1}^l h_{\varepsilon}(\alpha _r,R,N) \, ,
\]
where
\begin{displaymath}
h_{\varepsilon}(x,R,N)=\frac {\varepsilon \Delta ^{\varepsilon}}{2\Delta \sin \pi \beta ^2 }\left \{ \left [ \frac {e^{-i\pi \beta ^2(\frac {3}{2}-\varepsilon)-x}}{\sinh \left (x-\varepsilon \ln \frac {R}{N}+\frac {3i\pi \beta ^2}{2} \right )}- \frac {e^{-i\pi \beta ^2(\frac {1}{2}-\varepsilon)-x}}{\sinh \left (x-\varepsilon \ln \frac {R}{N}+\frac {i\pi \beta ^2}{2} \right )}\right ] + \right.
\end{displaymath}
\be
\left. + \Bigl [\beta ^2 \rightarrow -\beta ^2 \Bigr ]\right \} \, .
\label {h}
\ee
In the expression - see (\ref {sumf5}), (\ref {J}) - for the finite size correction of the energy what enters is the Fourier transform of (\ref {h}). Explicitly we have, using \cite {GR}:
\be
ik\hat h _{\varepsilon}(k,R,N)=2\varepsilon \pi k \frac {e^{-ik\varepsilon \ln \frac {R}{N}}}{\Delta \sin \pi \beta ^2}\, \frac {\sin \frac {\pi k \beta ^2}{2}}{\sin \frac {\pi k}{2}}\, \sinh \left (\pi k \beta ^2 -\frac {\pi k}{2}-\varepsilon i \pi \beta ^2\right )\, .
\ee
Therefore, the quantity $J_h(x)$ (\ref {J}) in this case is given by:
\[
J_h(x)=\lim _{N\rightarrow +\infty} \int _{-\infty}^{+\infty} \frac {dk}{2\pi}\, e^{ik\left \{ x-(1-\beta ^2)\varepsilon \ln  \left [\frac {A}{N}\sin \frac {\pi \beta ^2}{2(1-\beta ^2)} \right ]\right \} } \,
\frac {N \varepsilon \pi k \, \sinh \left (\pi k \beta ^2 -\frac {\pi k}{2}-\varepsilon i \pi \beta ^2\right )}{R \sin \pi \beta ^2 \, \cosh k \frac {\gamma}{2}} \, .
\]
This integral can be evaluated using the residue method. Since we are going to the limit $N\rightarrow +\infty$, we close the contour of integration in the upper $k$-complex half-plane if $\varepsilon =1$ and in the lower $k$-complex half-plane if $\varepsilon =-1$. In both cases the only non-vanishing contribution in the continuous limit is given by
the pole with smallest modulus, i.e. that at $k=i\varepsilon \pi/\gamma$, and hence turns out to be
\be
J_h(x)=-\frac {2A}{R}\, \frac {\sin \frac {\pi \beta ^2}{2(1-\beta ^2)} \, \sin \left ( \varepsilon \frac {\pi}{2} \frac {2\beta ^4-1}{1-\beta ^2}\right ) }{(1-\beta ^2)^2 \sin \pi \beta ^2} \, e^{-\varepsilon \frac {\pi}{\gamma}x} \, . \label {i}
\ee
Let us indicate by $E_{\tilde \omega}(R)$ the finite size correction to the energy of the $\tilde \omega$-vacuum, defined as the second term of the r.h.s. of (\ref {sumf5}) with $J_h$ given by (\ref {i}). The effective central charge is proportional to the $R$-indipendent quantity
\be
-\frac {6RE_{\tilde \omega}(R)}{\pi}=\frac {12A}{\pi}\, \frac {\sin \frac {\pi \beta ^2}{2(1-\beta ^2)} \, \sin \left ( \varepsilon \frac {\pi}{2} \frac {2\beta ^4-1}{1-\beta ^2}\right ) }{(1-\beta ^2)^2 \sin \pi \beta ^2}  
\int _{-\infty}^{+\infty}\frac {dx}{\pi} \, e^{-\varepsilon \frac {\pi}{\gamma}x} \,
{\mbox {Im}}\ln \left [1+e^{iZ(x+i0)}\right ] \, . \label{ecch}
\ee

As well explored \cite {DDV,KP,DDV1,FMQR}, the integral in (\ref {ecch}) can be computed using {\it a derivative lemma}, yielding the effective central charge
\be
c_{eff}= \left (1-\frac {6\beta ^2 \tilde \omega ^2}{\pi ^2}\right )
\, . \label{ceff}
\ee
Now we can identify the (conformal field) theory, i.e. the central charge $c(\beta ^2)$ independent on the twist $\tilde \omega$, and the conformal weight of the Bethe state $h$ reinterpreting (\ref {ceff}) as \cite {BLC}
\be
c_{eff}=c-24h \, , \label{ceff1}
\ee
where
\be
c=13-6\left (\beta ^2+\frac {1}{\beta ^2}\right ) \, \label{ceff2}
\ee
and
\be
h=\frac {\beta ^2 {\tilde \omega}^2}{4\pi ^2}+\frac {c-1}{24}=\frac {\omega ^2}{4\beta ^2}+\frac {c-1}{24} \, . \label{ceff3}
\ee
Nevertheless, if we want to implement constraint (\ref {omega}), $\omega$ can take only the values:
\be
\omega=\frac {s}{p^\prime} \, , \quad 0\leq s \leq p^\prime -1 \, . \label{omega1}
\ee
Therefore, inserting (\ref {omega1}) in (\ref {ceff}) we get the effective central charge for fixed $\beta ^2=p/p^\prime$:
\be
c_{eff}=\left (1-\frac {6s^2}{pp^\prime} \right) \, . \label{ceff4}
\ee
In the particular case of minimal unitary models, $p^\prime=p+1$, the effective central charge reads
\be
c_{eff}=\left (1-\frac {6s^2}{p(p+1)} \right) \, , \quad 0\leq s \leq p \, . \label{ceff5}
\ee
These values of $c_{eff}$ identify all primary states of minimal unitary models which are on the diagonal of the Kac table. Sublimits different from (\ref {par1}) may yield other primary states and the entire table should be completed by holes and/or complex Bethe's roots excitations \cite{Klu}.

\section {Continuous limit of the eigenvalues of the transfer matrix}
\setcounter{equation}{0}

It is surely interesting to use the results of section 3 in order to evaluate the continuous limit of the vacuum eigenvalue of the transfer matrix (\ref {tra}), in particular as in this limit the integrals of motion it generates in the asymptotic expansion are connected to the local conserved charges of the aforementioned conformal field theory \cite {BLZ,BLZ2}.

Referring for notations to formul{\ae} (\ref {tra}-\ref {rho}), we define
\be
F^+_N(\alpha )\equiv\ln {\Lambda}^+_N(\alpha ) \quad ,\quad F^-_N(\alpha )\equiv\ln { \Lambda}^-_N(\alpha ) \, , \label {tra2}
\ee
and we consider first $F^+$. From (\ref {tra1}, \ref {rho}) we have
\ba
&&F^+_N(\alpha )=\sum _{r=1}^{l}\Bigl [ \ln \frac {\sinh (\alpha -\alpha _r +i\pi \beta ^2)}{\sinh (\alpha -\alpha _r)}  -i \varepsilon \pi \beta ^2 \Bigr]+\nonumber \\
&+&\frac {N}{2}\left [ \ln \sinh \left ( \varepsilon  \alpha + \varepsilon \frac {i\pi \beta ^2}{2} -\ln \frac {R}{N} \right ) - \varepsilon \frac {i\pi \beta ^2}{2} +\ln \frac {2R}{N}-\varepsilon \alpha \right ]  \, . \label {tra3}
\ea
We want to evaluate for the vacuum
\begin{equation}
F^+(\alpha)\equiv\lim _{\stackrel {N\rightarrow +\infty}{N\Delta =R}} F^+_N(\alpha +\Lambda _N) \, .
\label{Fcont}
\end{equation}
The last addendum in (\ref {tra3}) gives a contribution whose continuous limit is zero if $0<\beta ^2 <1/2$ and infinity if $1/2<\beta ^2 <1$. However, one can get rid of this divergence by defining $F^+(\alpha)$ for $1/2<\beta ^2 <1$ as the analytic continuation of the same function for $0<\beta ^2 <1/2$.

With this regularisation procedure in mind, we will calculate, for $0<\beta ^2 <1/2$,
\be
F^+(\alpha)=\lim _{\stackrel {N\rightarrow +\infty}{N\Delta =R}}\left \{ \sum _{r=1}^{l}\left [ \ln \frac {\sinh (\alpha +\Lambda _N -\alpha _r +i\pi \beta ^2)}{\sinh (\alpha +\Lambda _N -\alpha _r)} -i\varepsilon \pi \beta ^2 \right ]\right \}
 \, , \label {tra4}
\ee
and we will release the constraint on $\beta ^2$. As already pointed out,
the constant term in (\ref {tra4}) has to be treated indipendently. Its contribution is $-i\varepsilon \pi \beta ^2 l $ and, since $l$ is even, we can restrict ourselves to the subsequences in which the integers $n$ and $\kappa$, contained in (\ref {par1}), are both even. With this restriction, using the relations (\ref {beta}), (\ref {par1}), the constant term in (\ref {tra4}) can be written as follows
\be
-i\varepsilon \pi \frac {p}{p^\prime} \kappa + 2\pi i {\Bbb Z}\, . \label {xx}
\ee
The multiples of $2\pi i $ can be discarded since the transfer matrix depends only on exp$\, [F^+]$. Remembering the definition of the twist (\ref {omega}),  contribution (\ref {xx}) becomes (the double square brackets denote the integer part):
\[
-i\varepsilon \pi \omega -i\varepsilon \pi  \left [ \left [ \frac {pk}{p^\prime} \right ] \right ]
\, .
\]
The integer part gives a factor to exp$\, [F^+]$ which can be equal to $\pm 1$, depending on the choice of $\kappa$. For the sake of simplicity we will carry on the calculations supposing that $\kappa$ is chosen in such a way that this factor is $1$: the vacuum eigenvalue of the transfer matrix for the other values of $\kappa$ differs form the result we will obtain in this section by a global factor $-1$ and hence its treatment is redundant. Therefore, without losing generality, we can conclude that the constant term in (\ref {tra4}) gives a contribution to $F^+$ which is equal, up to multiples of $2\pi i $, to $-i\varepsilon \pi \omega$.

Using this result and (\ref {sumf5}) for dealing with (\ref {tra4}), we obtain
\be
{F}^+(\alpha)={f}_b^+(\alpha )+\int _{-\infty}^{+\infty}\frac {dx}{\pi}
\Bigl [\int _{-\infty}^{+\infty}\frac {dk}{2\pi}e^{ikx} ik \hat f^+(k,\alpha) \hat J(k) \Bigr ]{\mbox {Im}}\ln \left [1+e^{i Z(x+i0)}\right ]
-i\varepsilon \pi \omega \, , \label {sumf6}
\ee
where ${f}_b^+$ is a bulk contribution, $\hat J(k)$ is given by (\ref {J}) and
\be
f^+(x,\alpha )=\ln \frac {\sinh \left (\alpha -x+i\pi \beta ^2 \right)}{\sinh (\alpha -x)}   \, . \label {f1}
\ee

After defining
\be
y=\alpha -i\frac {\gamma}{2} \, , \label {ydef}
\ee
we will study ${F}^+(y)={F}^+(\alpha)$ in the two regions:
\ba
{\mbox {REGION}} \, \, 1 \, \, &:& -\frac {\gamma}{2}< {\mbox {Im}}y < \frac {\gamma}{2} \label {reg1}\, ,  \\
{\mbox {REGION}} \, \, 2 \, \, &:& \frac {\gamma}{2}-\pi< {\mbox {Im}}y < -\frac {\gamma}{2} \label {reg2} \, .
\ea
The behaviour of ${F}^+(\alpha)$ in the other regions of the complex plane can be obtained using the periodicity property,
\be
F^+(\alpha +i\pi)=F^+(\alpha) \, , \label {perio}
\ee
which comes from expression (\ref {tra4}) for $F^+(\alpha)$.

Let us start with region 1. We remark that, when $|{\mbox {Im}}y|<\frac {\gamma}{2}$,
\be
f^+(x,\alpha)=-i \phi \left (x-y, \frac {\gamma}{2}\right)\, , \label {f3}
\ee
where $\phi$ is defined in (\ref {funz}). From (\ref {sumf6}) it follows that we have to calculate the Fourier transform of the derivative of $f^+(x,\alpha)$ as function of $x$. From the definition of $K(x)$ - first of the (\ref {not}) - and from its Fourier transform - second of the (\ref {four1}) - we obtain:
\be
ik\hat f^+ (k, \alpha)=-2\pi i e^{-iky}\frac {\sinh k\left (\frac {\pi}{2}-\frac {\gamma}{2}\right)}{\sinh k\frac {\pi}{2}} \, . \label {hatf1}
\ee
This result implies that, in this case, $J_{f^+}(x,\alpha)$ (\ref {J}) bears the form (after some simplifications)
\be
J_{f^+}(x-y)=\frac {1}{2i} \int _{-\infty}^{+\infty} {dk} e^{ik(x-y)}\frac {1}{\cosh k \frac {\gamma}{2}} \, . \label {I4}
\ee
This integral is finite if $|{\mbox {Im}}y|<\frac {\gamma}{2}$ and the result is \cite {GR}
\be
J_{f^+}(x-y)=\frac {\pi}{i\gamma}\frac {1}{\cosh \frac {\pi}{\gamma}(x-y)}
\, . \label {I5}
\ee
Now, after inserting (\ref {I5}) in (\ref {sumf6}), we have
\be
{F}^+(\alpha)={f}_b^+(\alpha)+\int _{-\infty}^{+\infty}\frac {dx}{i\gamma}\frac {1}{\cosh \frac {\pi}{\gamma}(x-y)}{\mbox {Im}}\ln \left [1+e^{iZ(x+i0)}\right ] -i\varepsilon \pi \omega \, ,
\label {sumf7}
\ee
and we need to calculate (\ref {bulk})
\be
{f}_b^+(\alpha)=\int _{-\infty}^{+\infty}\frac {dx}{2\pi} \frac {df^+(x,\alpha)}{dx}
\left [ A \varepsilon e^{- \varepsilon \frac {\pi}{\gamma}x }
-\varepsilon \tilde \omega \right ] \, . \label {bulk2}
\ee
Using (\ref {f3}) we arrive, after some manipulations, to the integral
\be
{f}_b^+(\alpha)=-\int _{-\infty}^{+\infty}\frac {dx}{2\pi}
\frac {\sinh i\pi \beta ^2}{\cosh \left (y-x-\frac {i\pi \beta ^2}{2}\right)\cosh \left(y-x+\frac {i\pi \beta ^2}{2} \right )}\left [ A \varepsilon e^{- \varepsilon \frac {\pi}{\gamma}x }
-\varepsilon \tilde \omega  \right ] \, , \label {bulk3}
\ee
which is simple to be explicitly calculated when $0<\beta ^2 < 1/2$:
\be
{f}_b^+(\alpha)=i\varepsilon A \, {\mbox {ctg}}\,  \frac {\pi ^2}{2\gamma}\, \,  e^{-\varepsilon \frac {\pi}{\gamma}y}+i\varepsilon \pi \omega
 \, , \label {bulk4}
\ee
and, instead, diverges if $1/2<\beta ^2 <1$. However, in the regularisation procedure we chose, we define the limit (\ref {Fcont}) for $1/2<\beta ^2 <1$ as the analytic continuation of its value for $0<\beta ^2 <1/2$.
Hence, collecting formul{\ae} (\ref {sumf7}) and (\ref {bulk4}), we get
\be
{F}^+(\alpha)=i\varepsilon A \, {\mbox {ctg}}\, \frac {\pi^2}{2\gamma}\, \,  e^{-\varepsilon \frac {\pi}{\gamma}y} +\int _{-\infty}^{+\infty}\frac {dx}{i\gamma}\frac {1}{\cosh \frac {\pi}{\gamma}(x-y)}{\mbox {Im}}\ln \left [1+e^{iZ(x+i0)}\right ] \, ,
\label {sumf8}
\ee
for the whole interval $0<\beta ^2 < 1$ and if $|{\mbox {Im}}y|<\frac {\gamma}{2}$.

\medskip

Let us now focus on region 2. The evaluation of $ik\hat f^+ (k, \alpha)$ can be done starting from (\ref {f1}) and using \cite {GR}:
\be
ik\hat f^+ (k, \alpha)=2\pi i e^{-ik\left (y+i\frac {\pi}{2}\right)}\frac {\sinh k\frac {\gamma}{2}}{\sinh k\frac {\pi}{2}} \, . \label {rehatf1}
\ee
It follows that $J_{f^+}(x,\alpha)$ is now given by the expression
\be
J_{f^+}(x,\alpha)=\frac {i}{2} \int _{-\infty}^{+\infty} {dk} e^{ik\left (x-y-i\frac {\pi}{2}\right)}\frac {\sinh k\frac {\gamma}{2}}{\cosh k \frac {\gamma}{2}\, \sinh k \left (\frac {\pi}{2}-\frac {\gamma}{2}\right ) } \, , \label {reI4}
\ee
which, as far as we know, cannot be expressed through simple operations on elementary functions.
On the other hand, the evaluation of the bulk term, defined by formula (\ref {bulk2}), can be performed explicitly. With the help of \cite {GR} we obtain the following result, valid in region 2 and again for
$0<\beta ^2 <1/2$:
\be
{f}_b^+(\alpha)=i\varepsilon A\, \frac {1}{\sin \frac {\pi ^2}{2\gamma}}\, e^{-\varepsilon \frac {\pi}{\gamma}\left (y+i\frac {\pi}{2}\right ) }+i\varepsilon \pi \omega-i\varepsilon \tilde \omega
 \, . \label {rebulk4}
\ee
With the regularisation procedure we discussed before, we can extend this result to the whole range of values $0<\beta ^2 <1$.
Putting together (\ref {reI4}), (\ref {rebulk4}) with (\ref {sumf6}), we finally obtain the expression for the value of ${F}^+(\alpha)$ in the entire interval $0<\beta ^2 <1$ and in the region 2, $\frac {\gamma}{2}-\pi< {\mbox {Im}}y < -\frac {\gamma}{2}$:
\ba
&&{F}^+(\alpha)=i\varepsilon A \, \frac {1}{\sin \frac {\pi^2}{2\gamma}}\, e^{-\varepsilon \frac {\pi}{\gamma}\left (y+i\frac {\pi}{2}\right )}-i\varepsilon \tilde \omega +\label {resumf8} \\
&+&\int _{-\infty}^{+\infty}\frac {dx}{\pi }\int _{-\infty}^{+\infty} {dk}e^{ik\left (x-y-i\frac {\pi}{2}\right )} \frac {i\sinh k \frac {\gamma}{2}}{2\cosh k\frac {\gamma}{2}\, \sinh k\left (\frac {\pi}{2}-\frac {\gamma}{2}\right)}{\mbox {Im}}\ln \left [1+e^{iZ(x+i0)}\right ] \, .
\nonumber
\ea
This result can be also obtained as analytical continuation of the function (\ref {sumf8}) beyond the singularity of the integral kernel at ${\mbox {Im}}y=-\gamma/2$: in the $x$-plane the pole crosses the real axis and we need to take it into account through the residue theorem and hence using equation (\ref {ddv}).

\medskip

Let us now consider $F^-$. From (\ref {tra1}), (\ref {rho}) and (\ref {tra2}) we have that
\ba
&&F^-_N(\alpha )=\sum _{r=1}^{l}\Bigl [ \ln \frac {\sinh (\alpha -\alpha _r -i\pi \beta ^2)}{\sinh (\alpha -\alpha _r)}  +i \varepsilon \pi \beta ^2 \Bigr]+\nonumber \\
&+&\frac {N}{2}\left [ \ln \sinh \left ( \varepsilon  \alpha - \varepsilon \frac {i\pi \beta ^2}{2} -\ln \frac {R}{N} \right ) +\ln \frac {2R}{N}-\varepsilon \alpha +\varepsilon \frac {i\pi \beta ^2}{2} \right ]  \, . \label {tra5}
\ea
We want to evaluate (for the vacuum)
\begin{equation}
F^-(\alpha)\equiv\lim _{\stackrel {N\rightarrow +\infty}{N\Delta =R}} F^-_N(\alpha +\Lambda _N) \, .
\label{Fcont-}
\end{equation}
Again, we have that the continuous limit of the last addendum in (\ref {tra5}) is zero if $0< \beta ^2 < 1/2$ and infinity if $1/2<\beta ^2 <1$. However, according to our regularisation procedure, we define $F^-$ for $1/2<\beta ^2 <1$ as the analytic continuation of its value for $0< \beta ^2 < 1/2$.
Hence, we have
\be
F^-(\alpha)=\lim _{\stackrel {N\rightarrow +\infty}{N\Delta =R}} \left \{ \sum _{r=1}^{l}\left [ \ln \frac {\sinh (\alpha +\Lambda _N-\alpha _r -i\pi \beta ^2)}{\sinh (\alpha +\Lambda _N -\alpha _r)}+\varepsilon i \pi \beta ^2 \right ] \right \}
 \, . \label {2tra4}
\ee
Comparing (\ref {2tra4}) with (\ref {tra4}), we notice that
\be
{F}^-(\alpha)=- {F}^+\left (\alpha-i\pi \beta ^2 \right ) \label{corr}
\, ,
\ee
and this means that the behaviour of $F^-(\alpha)$ in all the complex plane can be obtained from results (\ref {sumf8}), (\ref {resumf8}) and (\ref {perio}) for $F^+(\alpha)$. We stress that, because of the periodicity property (\ref {perio}), relation (\ref {corr}) can be alternatively written as
\be
{F}^-(\alpha)=- {F}^+\left (\alpha+i\gamma \right ) \label{corr2}
\, .
\ee
To summarise, we have shown that the vacuum eigenvalue of the transfer matrix,
\be
\Lambda (\alpha)={\mbox {exp}}[F^+(\alpha)]+{\mbox {exp}}[-F^+(\alpha+i\gamma)] \, , \label{xxx}
\ee
can be obtained from relations (\ref {sumf8}), (\ref {resumf8}) and from the periodicity property (\ref {perio}).

\subsection {Asymptotic behaviour}

We now want to extract information on the vacuum eigenvalues of the local conserved charges of conformal field theories. In the classical case ($\beta ^2 \rightarrow 0$) these appear \cite {FS} as coefficients of the asymptotic expansion of the logarithm of the transfer matrix ${\Lambda} (\alpha)$ around the complex $\lambda =e^{\alpha}=\infty$. Likewise to \cite {FS} we expect that the local integrals of motion are encoded in $F^+(\alpha)$. But now there are two different asymptotic expansions of
$F^+(\alpha)$ following from (\ref {sumf8}) and (\ref {resumf8}) respectively. More explicitly, since
\[
{\mbox {Re}}  y \rightarrow -\varepsilon \infty \Rightarrow \frac {1}{\cosh \frac {\pi}{\gamma}(x-y)}\, \, \, {\stackrel {\cdot}{=}}\, \, \, 2 \sum _{n=0}^{+\infty} (-1)^n e^{-\varepsilon \frac {\pi}{\gamma}(x-y)(2n+1)} \, ,
\]
where the symbol $ {\stackrel {\cdot}{=}}$ represents asymptotic expansion,
from (\ref {sumf8}) we have that, when ${\mbox {Re}} y \rightarrow -\varepsilon \infty$ in the strip $|{\mbox {Im}}y|<\frac {\gamma}{2}$,
\ba
F^+(\alpha) &{\stackrel {\cdot}{=}}&i\varepsilon A \, {\mbox {ctg}} \, \frac {\pi ^2}{2\gamma}\, \, e^{-\varepsilon \frac {\pi}{\gamma}y}+ \label{expans8}
\\
&+&\frac {2}{i\gamma}\sum _{n=0}^{+\infty}(-1)^n \, e^{\varepsilon\frac {\pi}{\gamma}y(2n+1)}
\int _{-\infty}^{+\infty} dx \, e^{-\varepsilon \frac {\pi}{\gamma}x(2n+1)}{\mbox {Im}}\ln \left [1+e^{iZ(x+i0)}\right ] \, ,
\nonumber
\ea
whereas, from (\ref {resumf8}), it follows that, when ${\mbox {Re}}  y \rightarrow -\varepsilon \infty$ in the strip $\frac {\gamma}{2}-\pi< {\mbox {Im}}y < -\frac {\gamma}{2}$,
\ba
&&F^+(\alpha)\, \, \, {\stackrel {\cdot}{=}} \, -i\varepsilon \tilde \omega+i \varepsilon A \, \frac {1}{\sin \frac {\pi ^2}{2\gamma}}\, e^{-\varepsilon \frac {\pi}{\gamma}\left (y+i\frac {\pi}{2}\right)}+ \label{expans5}
\\
&+&{2i}\sum _{n=0}^{+\infty}\left \{ \frac {(-1)^{n+1}}{\gamma \cos \frac {\pi ^2(2n+1)}{2\gamma}}\,  e^{\varepsilon \frac {\pi}{\gamma}\left (y+i\frac {\pi}{2}\right)(2n+1)}
\int _{-\infty}^{+\infty} dx \, e^{-\varepsilon \frac {\pi}{\gamma}x(2n+1)}{\mbox {Im}}\ln \left [1+e^{iZ(x+i0)}\right ] + \right. \nonumber \\
&+&\left. \frac {(-1)^{n}\tan \frac {\pi \gamma (n+1)}{\pi-\gamma}}{\pi-\gamma} \, e^{\varepsilon \frac {2\pi}{\pi-\gamma}\left (y+i\frac {\pi}{2}\right)(n+1)}
\int _{-\infty}^{+\infty} dx \, e^{-\varepsilon \frac {2\pi}{\pi-\gamma}x(n+1)}{\mbox {Im}}\ln \left [1+e^{iZ(x+i0)}\right ] \right \} \, . \nonumber 
\ea
In fact, likewise to \cite {FS}, our final aim is to expand asymptotically the logarithm of the transfer matrix
\be
\ln {\Lambda} (\alpha)=\ln \left \{ {\mbox {exp}}[F^+(\alpha)]+{\mbox {exp}}[-F^+(\alpha+i\gamma)] \right \} \, , \label{Trans}
\ee
for $y \in {\Bbb C}$ and
\be
{\mbox {Re}} y \rightarrow -\varepsilon \infty  \, ,
\ee
with the parameter $\beta ^2$ restricted to the {\it semiclassical interval}
\be
0<\beta ^2 <\frac {1}{2} \, .
\ee
We will carry on the asymptotic analysis of $\ln {\Lambda} (\alpha)$ dividing the one-period strip
\be
\frac {\pi}{2}\beta ^2 -\pi <{\mbox {Im}} y < \frac {\pi}{2}\beta ^2
\ee
in the following domains:
\ba
{\mbox {DOMAIN}} \, \, 1 \, \, &:& \frac {\pi}{2}\beta ^2 -\pi < {\mbox {Im}}y < \frac {\gamma}{2} -\pi \, , \label {dom1} \\
{\mbox {DOMAIN}} \, \, 2 \, \, &:& \frac {\gamma}{2}-\pi< {\mbox {Im}}y < -\frac {\gamma}{2} \, , \label {dom2} \\
{\mbox {DOMAIN}} \, \, 3 \, \, &:& -\frac {\gamma}{2} < {\mbox {Im}}y < \pi -\frac {3}{2} \gamma \, , \label {dom3} \\
{\mbox {DOMAIN}} \, \, 4 \, \, &:& \pi-\frac {3}{2}\gamma< {\mbox {Im}}y < \frac {\pi}{2}\beta ^2 \, . \label {dom4}
\ea

Let us start from domain 1. The expansion of ${F}^+(\alpha)$ is given by (\ref {expans8}) with $y\rightarrow y+i\pi$, thanks to the periodicity (\ref {perio}), and similarly the asymptotics of $-{F}^+(\alpha+i\gamma)$ is equal to (\ref {expans8}) with a global minus sign and with $y\rightarrow y+i\gamma$. This shift compensates the global minus sign, and therefore we have 
\ba
F^+(\alpha)&{\stackrel {\cdot}{=}}& i \varepsilon A \, {\mbox {ctg}} \, \frac {\pi ^2}{2\gamma}\, \, e^{-\varepsilon \frac {\pi}{\gamma}(y+i\pi)}+ \label{expans2}
\\
&+&\frac {2}{i\gamma}\sum _{n=0}^{+\infty}(-1)^n \, e^{\varepsilon\frac {\pi}{\gamma}(y+i\pi)(2n+1)}
\int _{-\infty}^{+\infty} dx \, e^{-\varepsilon\frac {\pi}{\gamma}x(2n+1)}{\mbox {Im}}\ln \left [1+e^{iZ(x+i0)}\right ] \, ,
\nonumber
\ea
\ba
-F^+(\alpha+i\gamma)&{\stackrel {\cdot}{=}} & i\varepsilon A \, {\mbox {ctg}} \, \frac {\pi ^2}{2\gamma}\, \, e^{-\varepsilon \frac {\pi}{\gamma}y}+ \label{expans3}
\\
&+&\frac {2}{i\gamma}\sum _{n=0}^{+\infty}(-1)^n \, e^{\varepsilon \frac {\pi}{\gamma}y(2n+1)}
\int _{-\infty}^{+\infty} dx \, e^{-\varepsilon \frac {\pi}{\gamma}x(2n+1)}{\mbox {Im}}\ln \left [1+e^{iZ(x+i0)}\right ] \, .
\nonumber
\ea
Which of these two expansions dominates in the asymptotics of the logarithm of the transfer matrix (\ref {Trans}) is given by the comparison of the asymptotic values of their leading terms. Hence,
\[
\lim _{{\rm {Re}} y \rightarrow -\varepsilon \infty}{\mbox {Re}} \left [
i \varepsilon A \, {\mbox {ctg}} \, \frac {\pi ^2}{2\gamma}\, \, e^{-\varepsilon \frac {\pi}{\gamma}y}
 -i \varepsilon A \, {\mbox {ctg}} \, \frac {\pi ^2}{2\gamma}\, \, e^{-\varepsilon \frac {\pi}{\gamma}(y+i\pi)}
\right ] = +\infty \,
\]
implies that (\ref {expans3}) dominates:
\ba
&&\ln {\Lambda} (\alpha)\, \, \, {\stackrel {\cdot}{=}}\, \,  -F^+(\alpha+i\gamma) \, \, \, {\stackrel {\cdot}{=}}\,  \label{expans4}
\\
&&i \varepsilon A \, {\mbox {ctg}} \, \frac {\pi ^2}{2\gamma}\, \, e^{-\varepsilon \frac {\pi}{\gamma}y}+\frac {2}{i\gamma}\sum _{n=0}^{+\infty}(-1)^n \, e^{\varepsilon\frac {\pi}{\gamma}y(2n+1)}
\int _{-\infty}^{+\infty} dx \, e^{-\varepsilon \frac {\pi}{\gamma}x(2n+1)}{\mbox {Im}}\ln \left [1+e^{iZ(x+i0)}\right ] \, .
\nonumber
\ea

\medskip

In the domain 2, when ${\mbox {Re}} y \rightarrow -\varepsilon \infty$ the asymptotic expansion of $F^+(\alpha)$ is given by formula (\ref {expans5}), whereas the asymptotic expansion of $-F^+(\alpha +i\gamma)$ is still given by relation (\ref {expans3}). The leading terms of (\ref {expans3}) and (\ref {expans5}) satisfy the limit
\[
\lim _{{\rm {Re}} y \rightarrow -\varepsilon \infty}{\mbox {Re}} \left [
i\varepsilon A \, {\mbox {ctg}} \, \frac {\pi ^2}{2\gamma}\, \, e^{-\varepsilon \frac {\pi}{\gamma}y}-
i\varepsilon A \, \frac {1}{\sin \frac {\pi ^2}{2\gamma}}\, e^{-\varepsilon \frac {\pi}{\gamma}\left (y+i\frac {\pi}{2}\right)}
\right ] = +\infty \, ,
\]
therefore, when ${\mbox {Re}} y \rightarrow -\varepsilon \infty$, $\ln {\Lambda} (\alpha)\, \, {\stackrel {\cdot}{=}}\, \, -F^+(\alpha+i\gamma)$, i.e. in domain 2 the expansion of the logarithm of the transfer matrix is still given by formula (\ref {expans4}).

\medskip

We now analyse domain 3. In this domain the asymptotic expansion of $F^+(\alpha)$ is given by expression (\ref {expans8}), whereas the asymptotic expansion of $-F^+(\alpha+i\gamma)$ is given by minus expression (\ref {expans5}) in which $y$ is substituted by $y+i\gamma-i\pi$. Hence, up to the leading term
\be
F^+(\alpha) \simeq i \varepsilon A \, {\mbox {ctg}} \, \frac {\pi ^2}{2\gamma}\, \, e^{-\varepsilon \frac {\pi}{\gamma}y} \label {expans6}
\ee
and, on the other hand,
\be
-F^+(\alpha+i\gamma)\simeq  -i \varepsilon A \, \frac {1}{\sin \frac {\pi ^2}{2\gamma}}\, e^{-\varepsilon \frac {\pi}{\gamma}\left (y-i\frac {\pi}{2}+i\gamma\right)} \, .
\ee
Again, one can show that in the limit ${\mbox {Re}} y \rightarrow -\varepsilon \infty$ the leading term (\ref {expans6}) is dominating:
\[
\lim _{{\rm {Re}} y \rightarrow -\varepsilon \infty}{\mbox {Re}} \left [
i \varepsilon A \, {\mbox {ctg}} \, \frac {\pi ^2}{2\gamma}\, \, e^{-\varepsilon \frac {\pi}{\gamma}y}+
i \varepsilon A \, \frac {1}{\sin \frac {\pi ^2}{2\gamma}}\, e^{-\varepsilon \frac {\pi}{\gamma}\left (y-i\frac {\pi}{2}+i\gamma\right)}
\right ] = +\infty \, ;
\]
therefore, when ${\mbox {Re}} y \rightarrow -\varepsilon \infty$, $\ln {\Lambda} (\alpha)\, \, \, {\stackrel {\cdot}{=}}\, \, \, F^+(\alpha)$ (\ref {expans8}) and, since (\ref {expans8}) coincides with (\ref {expans3}) and (\ref {expans4}), the asymptotic expansion of $\ln {\Lambda}(\alpha)$ in the domain 3 is still given by (\ref {expans4}).

\medskip

We are now left with domain 4. Again, the asymptotics of $F^+(\alpha)$ is given by expression (\ref {expans8}), and, as a consequence, its leading term by (\ref {expans6}). On the other hand, $-F^+(\alpha+i\gamma)$ is given by minus expression (\ref {expans8}) in which $y$ is substituted by $y+i\gamma-i\pi$ and, consequently, its leading term
\be
-F^+(\alpha+i\gamma)\simeq -i \varepsilon A \, {\mbox {ctg}} \, \frac {\pi ^2}{2\gamma}\, \, e^{-\varepsilon \frac {\pi}{\gamma}
(y+i\gamma-i\pi)} \,
\ee
is subdominant, i.e.
\[
\lim _{{\rm {Re}} y \rightarrow -\varepsilon \infty}{\mbox {Re}}
\left [
i\varepsilon A \, {\mbox {ctg}} \, \frac {\pi ^2}{2\gamma}\, \, e^{-\varepsilon \frac {\pi}{\gamma}y}+
i\varepsilon A \,{\mbox {ctg}} \, \frac {\pi ^2}{2\gamma}\, \,  e^{-\varepsilon \frac {\pi}{\gamma}\left (y+i\gamma-i\pi\right)}
\right ] = +\infty \, .
\]
Therefore, when ${\mbox {Re}}  y \rightarrow -\varepsilon \infty$, the asymptotic expansion of $\ln {\Lambda} (\alpha)$ is still given by (\ref {expans4}).

\medskip

The validity of such a picture can be partially extended to the interval
\be
\frac {1}{2}<\beta ^2 <1 \, .
\ee
Indeed, following the same procedure as in the semiclassical case, one can show that in the strip
\[
-\frac {3}{2}\gamma <{\mbox {Im}} y < \frac {\gamma}{2}
\]
the asymptotic expansion of $\ln {\Lambda} (\alpha)$ is given by (\ref {expans4}); however, in the strip
\[
 \frac {\gamma}{2} -\pi <{\mbox {Im}} y < -\frac {3}{2}\gamma \, ,
\]
which completes the $i\pi$ periodicity, the asymptotic development of $\ln {\Lambda} (\alpha)$ does not follow in a simple way from (\ref {expans8}, \ref {expans5}). However, it is still possible to write such an expression, which is found to be different from (\ref {expans4}).

\medskip

To summarise, we have shown that in the limit ${\mbox {Re}} y \rightarrow -\varepsilon \infty$ and in the domain $\frac {\pi}{2}\beta ^2 -\pi <{\mbox {Im}}y< \frac {\pi}{2} \beta ^2 $ the asymptotic expansion of the logarithm of the vacuum eigenvalue of the transfer matrix in the semiclassical domain $0<\beta ^2 <1/2$ is given by
\be
\ln {\Lambda}(\alpha)\, \, \, {\stackrel {\cdot}{=}}\, \, \,  -i \varepsilon m \, e^{-\varepsilon \frac {\pi}{\gamma}y}+
\frac {2}{i\pi}\sum _{n=0}^{+\infty}(-1)^n e^{\varepsilon \frac {\pi}{\gamma}y(2n+1)}
c_{n+1}I_{2n+1} \, ,
\label{expans15}
\ee
where
\be
m=-A \,  {\mbox {ctg}} \, \frac {\pi ^2}{2\gamma} \, , \label{m}
\ee
\be
c_{n+1}=-\frac {1+\xi}{(n+1)!} \frac {\beta ^{2n+2}\pi ^{\frac {3}{2}}}{2} \frac {\Gamma \left (\left (n+\frac {1}{2}\right )(1+\xi)\right )}{\Gamma \left (1+\left (n+\frac {1}{2}\right )\xi\right )} [\Gamma (1-\beta ^2)]^{-(2n+1)(1+\xi)} \,  ,  \quad \xi \equiv \frac {\beta ^2}{1-\beta ^2} \label{Cn1} \, ,
\ee
and
\be
I_{2n+1}=\frac {1}{c_{n+1}}\frac {\pi}{\gamma}\int _{-\infty}^{+\infty} dx \, e^{-\varepsilon \frac {\pi}{\gamma}x(2n+1)}{\mbox {Im}}\ln \left [1+e^{iZ(x+i0)}\right ] \, . \label{In}
\ee
Of course, the asymptotic behaviour of $\ln {\Lambda} (\alpha)$ in all the complex plane follows from (\ref {expans15}) and the periodicity property (\ref {perio}).
Moreover, relation (\ref {expans15}) is still valid in the domain $1/2<\beta ^2 <1$, provided we restrict it to the strip $-\frac {3}{2}\gamma <{\mbox {Im}} y < \frac {\gamma}{2}$. Without this restriction, we would have that the expansion is governed also by the dual non-local charges, appearing in the second
series of (\ref {expans5}).

\medskip

In the next subsection we are going to prove that the local integrals of motion $I_{2n+1}$ in (\ref {expans15}) are also normalised by the $c_n$ as usually in Conformal Field Theories. For the entire interval $0<\beta ^2 <1$ they are given by the asymptotic expansion (\ref {expans8}) of $F^+(\alpha)$. The case $n=0$
\[
I_1=\frac {1}{c_{1}}\frac {\pi}{\gamma}\int _{-\infty}^{+\infty} dx \, e^{-\varepsilon \frac {\pi}{\gamma}x}{\mbox {Im}}\ln \left [1+e^{iZ(x+i0)}\right ]
\]
can be evaluated using the derivative lemma \cite {DDV,DDV1}, but an explicit formula for the cases $n\geq 1$ is still missing.

Moreover, we want to stress that, when $\varepsilon =-1$, formul{\ae} (\ref {expans15}-\ref {In}) coincide with Conjecture (2.21) of \cite {BLZ2} for the regime $0<\beta ^2 <1/2$. In fact, the two asymptotic expansions for $\ln {\Lambda}(\alpha)$ and their domains of validity are the same if we compare the quantum Lax operator (\ref {lax}) and the classical one in \cite {FR} to obtain the connecting $\lambda^2= e^{2\alpha}=e^{2y+i\gamma}$, and if we notice that (\ref {m}) reproduces formula (4.15) of \cite {BLZ2},
\be
m=2{\sqrt {\pi}}\, \, \frac {\Gamma \left (\frac {1-2\beta ^2}{2-2\beta ^2} \right )}{\Gamma \left (\frac {2-3\beta ^2}{2-2\beta ^2} \right )}\Gamma (1-\beta ^2)^{\frac {1}{1-\beta ^2}} \, ,
\ee
after choosing $A$ as in Remark 4. On the other hand, we have found two different asymptotic expansions for $F^+(\alpha)$, i.e. formula (\ref {expans8}), valid for $-\gamma/2<{\mbox {Im}}y<\gamma/2$, which coincides with (4.14) of \cite {BLZ2}, and formula (\ref {expans5}), valid for $\gamma/2-\pi<{\mbox {Im}}y<-\gamma/2$.

\subsection{Limit ${\tilde \omega} \rightarrow +\infty$}

This limit {\it linearises} equation (\ref {ddv}) into an inhomogeneous integral Wiener-Hopf equation of second kind (which is solvable using standard Fourier transform techniques).
In fact, after introducing new normalisations
\be
\theta = \frac {\pi}{\gamma}x \, , \quad \tilde Z(\theta)=Z\left (\frac {\gamma}{\pi}\theta\right) \, , \quad \tilde G(\theta)=\frac {\gamma}{\pi}G
\left (\frac {\gamma}{\pi}\theta\right) \, ,
\ee
equation (\ref {ddv}) reads
\be
\tilde Z(\theta)=- A\varepsilon e^{- \varepsilon \theta}+
2\int_{-\infty}^{+\infty}{d\theta^\prime}\, \tilde G(\theta-\theta^\prime){\mbox {Im}}\ln \left [1+e^{i\tilde Z(\theta ^\prime+i0)}\right ]+\varepsilon \tilde \omega \, .
\label{ddv1}
\ee
Let us confine ourselves to the case $\varepsilon =-1$ (the case $\varepsilon =1$ can be treated along the same lines). We will show in the following that equation (\ref {ddv1}) is properly valid in the range $0\leq\tilde \omega < \pi/2\beta ^2 $ -- although the results of Section 3 hold for all $\tilde \omega \geq 0$ thanks to their analyticity. Therefore, we calculate the exact solution of (\ref {ddv1}) in the limit ${\tilde \omega} \rightarrow +\infty$, meaning by that the analytic continuation of the solution valid in $0\leq\tilde \omega < \pi/2\beta ^2 $. Although the periodic solution do exist, we prove that the analytic solution is not periodic, for we need to modify the (\ref{ddv1}) itself by imposing analytic continuation. Indeed, its derivation is based on the assumption that the only points $\theta ^{(r)}$ on the real axis for which $e^{i\tilde Z(\theta ^{(r)})}=-1$ are the Bethe roots and this is true only if $0\leq\tilde \omega < \pi/2\beta ^2 $: in this case, the {\it holes} $\theta ^{(h)}$, such that $e^{i\tilde Z(\theta ^{(h)})}=-1$ and do not satisfy the Bethe equations (\ref {Bethe}), all lie on the axis ${\mbox {Im}}\, \theta ^{(h)}=\pi$. We see this because these holes are also the zeroes of the transfer matrix (\ref {xxx}) and we also understand that when $\tilde \omega$ increases from $0$ to
$\pi/2\beta ^2$ the real part of the first hole from left decreases; at $\tilde \omega =\pi/2\beta ^2$, the real part of that hole reaches $-\infty$:
\be
\lim _{{\rm {Re}}\, \theta \rightarrow -\infty}\Lambda (\theta +i0)=0 \Rightarrow e^{i\beta ^2 \tilde \omega}+e^{-i\beta ^2 \tilde \omega}=0 \, ; \label{holmig}
\ee
then for $\tilde \omega >\pi/2\beta ^2$ this hole {\it jumps} onto the real axis and {\it migrates} to the right \footnote[1]{We are really indebted to R. Tateo who has suggested us the possibility of the presence of holes on the real axis, although our treatment would not change if the holes were not present. Yet their real presence is conceptually important.}.
We have borrowed our reasoning from a slightly different context \cite {DDT}. Nevertheless we can go further in our analysis. Indeed, it is known that the presence of holes on the real axis for $\tilde \omega > \pi/2\beta ^2$ implies that the integral term of (\ref {ddv1}) - which should count only contributions coming from the poles of ${\mbox {Im}}\ln \left [1+e^{i\tilde Z(\theta ^\prime+i0)}\right ]$ corresponding to Bethe roots - in fact receives contribution also from the holes on the real axis. Therefore, in the presence of $N_h$ holes on the real axis at the positions $\theta _1, \ldots \theta _{N_h}$, we need to subtract the holes contribution as follows \cite {FMQR, Klu}:
\be
\tilde Z(\theta)=A e^{\theta}+
2\int_{-\infty}^{+\infty}{d\theta^\prime}\, \tilde G(\theta-\theta^\prime){\mbox {Im}}\ln \left [1+e^{i\tilde Z(\theta ^\prime+i0)}\right ]-\tilde \omega-i \sum _{h=1}^{N_h}\ln S(\theta -\theta _h) \, .
\label{ddvh}
\ee
In (\ref {ddvh}) we have introduced the sine-Gordon soliton-soliton scattering amplitude $S(\theta)$, whose explicit expression is
\be
S(\theta)={\mbox {exp}}\left [ \frac {i}{2}\int _{-\infty}^{+\infty}dk \frac {\sin k\theta}{k} \frac {\sinh \left (\frac {k\pi ^2 }{2\gamma}-k\pi \right )}{\cosh \frac {k\pi}{2} \sinh \left (\frac {k\pi ^2 }{2\gamma}-\frac {k\pi}{2} \right )} \right ] \, , \label{Sdef}
\ee
and whose connection with $\tilde G(\theta)$ is
\[
\tilde G(\theta)=-\frac {i}{2\pi}\frac {d}{d\theta}\ln S(\theta) \, .
\]
Likewise, in the presence of holes the quantity $F^+(\theta)$ (\ref {sumf8}, \ref {resumf8}) is modified into
\be
{F^+}^{(h)}(\theta)=F^+(\theta)+2i\sum _{h=1}^{N_h}\arctan e^{(\theta _h -\theta)} \quad , \quad \theta =\frac {\pi}{\gamma}y \, . \label{modf}
\ee
As a result we can find the exact values of $\tilde \omega$ at which one additional hole migrates onto the real axis. Indeed, the condition
\[
\lim _{{\rm {Re}}\, \theta \rightarrow -\infty}\Lambda ^{(h)}(\theta +i0)=0 \, ,
\]
applied to the expression of the vacuum eigenvalue of the transfer matrix in the presence of holes on the real axis,
\be
\Lambda ^{(h)}(\theta)={\mbox {exp}}[{F^+}^{(h)}(\theta)]+{\mbox {exp}}[-{F^+}^{(h)}(\theta+i\pi)] \, , \label {Trahol}
\ee
is still equivalent to
\[
e^{i\beta ^2 \tilde \omega}+e^{-i\beta ^2 \tilde \omega}=0 \, ,
\]
because in the limit ${\rm {Re}}\, \theta \rightarrow -\infty$ the added sum in (\ref {modf}) produces in (\ref {Trahol}) an overall factor $e^{i\pi N_h}$. Therefore, we conclude that when $\left (n+\frac {1}{2}\right )\frac {\pi}{\beta ^2}< \tilde \omega < \left (n+\frac {3}{2}\right )\frac {\pi}{\beta ^2}$ we have that $n+1$ holes are moving right on the real axis.

\medskip

We are now going to write the nonlinear integral equation (\ref {ddvh}) and the eigenvalues (\ref {Trahol}) in an alternative form which is suitable for
the study of the limit $\tilde \omega \rightarrow +\infty$: the integration path can be shifted in order to avoid contributions from the incoming holes. In this respect, we make the hypothesis that there is a separator $s(\tilde \omega)$ on the real axis such that for all the Bethe roots $\theta ^{(r)}$ and for all the holes on the real axis $\theta ^{(h)}$
\be
\theta ^{(h)} \leq s(\tilde \omega) \leq \theta ^{(r)} \, .
\ee
Under this hypothesis we can move down the integration path $(-\infty +i0, s(\tilde \omega )+i0)$ in
\be
{\cal I} (\theta)=2\int_{-\infty}^{+\infty}{d\theta^\prime} \tilde G(\theta - \theta ^\prime) {\mbox {Im}}\ln \left [1+e^{i\tilde Z(\theta ^\prime+i0)}\right ] \, ,
\label{funF}
\ee
of (\ref {ddv1}), in absence of holes ($0\leq\tilde \omega <\pi/2\beta ^2$), in such a way that, when the holes reach the real axis, they will never cross the integration path:
\ba
&&{\cal I} (\theta)=-i\int_{-\infty}^{s(\tilde \omega)}{d\theta^\prime} \tilde G(\theta - \theta ^\prime) \ln \left [1+e^{i\tilde Z(\theta ^\prime-i0)}\right ] - \label{funF1} \\
&-&i\int_{s(\tilde \omega)}^{+\infty}{d\theta^\prime} \tilde G(\theta- \theta ^\prime) \ln \left [1+e^{i\tilde Z(\theta ^\prime+i0)}\right ]+
i\int_{-\infty}^{+\infty}{d\theta^\prime} \tilde G(\theta- \theta ^\prime) \ln \left [1+e^{-i\tilde Z(\theta ^\prime-i0)}\right ] \, . \nonumber
\ea
This modified expression contains only contributions from the Bethe roots and holds by construction for any $\tilde \omega$. Clearly, when $0\leq\tilde \omega <\pi/2\beta ^2$ it coincides with (\ref {funF}). Algebraic manipulations allow us to rewrite (\ref {funF1}) in the compact form
\be
{\cal I} (\theta)=\int_{-\infty}^{s(\tilde \omega)}{d\theta^\prime} \tilde G(\theta - \theta ^\prime) \tilde Z(\theta ^\prime) +2\int_{s(\tilde \omega)}^{+\infty}{d\theta^\prime} \tilde G(\theta - \theta ^\prime) {\mbox {Im}}\ln \left [1+e^{i\tilde Z(\theta ^\prime+i0)}\right ] \, , \label{funF2}
\ee
where the first integral contains $\tilde Z(\theta ^\prime)$ thanks to the absence of holes and roots. Eventually, using relation (\ref {funF2}) we are able to extend analytically the equation (\ref {ddv1}) $\forall \, \, \tilde \omega \geq 0$:
\ba
\tilde Z(\theta)&=&A e^{\theta}-\tilde \omega +\int _{-\infty}^{s(\tilde \omega)}
{d\theta^\prime}\, \tilde G(\theta-\theta^\prime) \tilde Z(\theta ^\prime) +\nonumber \\
&+&2\int_{s(\tilde \omega)}^{+\infty}{d\theta^\prime}\, \tilde G(\theta-\theta^\prime){\mbox {Im}}\ln \left [1+e^{i\tilde Z(\theta ^\prime+i0)}\right ] \, .
\label{ddv2}
\ea
This form is very suitable for the study of the limit $\tilde \omega \rightarrow +\infty$ because all the nonlinearity is confined in the last term.
In fact, the last term can be neglected in the limit $\tilde \omega \rightarrow +\infty$, for it can be approximated replacing $\tilde Z(\theta)$ with $A e^{\theta}-\tilde \omega$:
\be
I^{nl}(\theta)\simeq 2\int_{s(\tilde \omega)}^{+\infty}{d\theta^\prime}\, \tilde G(\theta-\theta^\prime){\mbox {Im}}\ln \left [1+e^{i(A e^{(\theta ^\prime +i0)}-\tilde \omega)}\right ] \, . \label{Inl}
\ee
After a simple change of variable we have
\be
I^{nl}(\theta)\simeq 2\int _{-\tilde \omega +Ae^{s(\tilde \omega)}}^{+\infty} dt \frac {\tilde G\left (\theta-\ln \frac {t+\tilde \omega}{A}\right)}{t+\tilde \omega}
{\mbox {Im}}\ln \left [1+e^{it(1+i0)}\right ] \, . \label{kyra}
\ee
Now, for $\theta<\theta ^\prime$ the integral (\ref {G(x)}) furnishes us an asymptotic expansion ($\theta-\theta ^\prime \rightarrow -\infty$) on residues \[
\tilde G(\theta -\theta ^\prime)= \sum _{i}c_i e^{a_i(\theta -\theta ^\prime)} \, , \quad a_i>0 \, .
\]
Using this expansion we easily argue that we can neglect the last term of (\ref {ddv2}) in the limit $\tilde \omega \rightarrow +\infty$ provided for instance the inferior integration limit of (\ref {kyra})
\be
-\tilde \omega +Ae^{s(\tilde \omega)} \rightarrow +\infty \, , \quad {\mbox {as $\tilde \omega \rightarrow +\infty$}} \, .
\ee
This is realised by the asymptotic behaviour - $A$ can be always chosen $>1$ -
\be
\tilde \omega \rightarrow +\infty \quad , \quad  s(\tilde \omega) \simeq \ln \tilde \omega \, , \label{limB}
\ee
which will be proved {\it a posteriori} - with the existence proof of a separator - using the explicit solution.

In conclusion, we can approximate in the limit $\tilde\omega \rightarrow +\infty$ equation (\ref {ddv2}) with a linear equation of Wiener-Hopf type,
\be
\tilde Z(\theta)=Ae^{\theta}-\tilde \omega+
\int_{-\infty}^{s(\tilde \omega)}{d\theta^\prime}\, \tilde G(\theta-\theta^\prime) \tilde Z(\theta ^\prime) \, .
\label{wh}
\ee
We can solve this equation with the usual factorisation technique \cite {MF} finding the Fourier transform of
\be
\tilde Z_-(\theta)\equiv \left \{ \begin{array} {ll}\tilde Z(\theta) & {\mbox {if $\theta <s(\tilde \omega)$}} \\ 0 & {\mbox {otherwise}} \end{array} \right. \, ,
\ee
obviously defined as
\be
{\hat Z}_- (k)=\int_{-\infty}^{s(\tilde \omega)}d\theta e^{-ik\theta}\tilde Z (\theta) \,  . \label{fk}
\ee
This integral is well defined when ${\mbox {Im}} k >0$ because, when $\theta \rightarrow -\infty$, $\tilde Z(\theta)$ goes to a non-zero constant.

With the choice of $A$ as in Remark 4 the solution of (\ref {wh}) can be expressed by
\be
{\hat Z}_-(k)=-\frac {2i\pi ^{\frac {3}{2}}}{k} \left (\frac {\beta ^2 \tilde \omega}{2\pi}\right )^{1-ik}\frac {\Gamma \left (1-ik\frac {1+\xi}{2}\right )}{\Gamma \left (\frac {3}{2}-\frac {ik}{2}\right) \Gamma \left (1-ik\frac {\xi}{2} \right)} \left [ \frac {\Gamma \left ( \frac {\xi}{2}\right ) \Gamma \left (\frac {1}{2}-\frac {\xi}{2} \right )}{{\sqrt \pi} M} \right ]^{-ik} \, . \label {fk1}
\ee
Inserting in (\ref {fk1}) the expression of $M$ (\ref {M}), we simplify the solution:
\be
{\hat Z}_-(k)=-\frac {i\, \pi ^{ik+\frac {1}{2}}\, 2^{ik}}{k \beta ^{2ik-2}
\, \tilde \omega ^{ik-1}} \frac {\Gamma \left (1-ik\frac {1+\xi}{2}\right )}{\Gamma \left (\frac {3}{2}-\frac {ik}{2}\right) \Gamma \left (1-ik\frac {\xi}{2} \right)} \, [\Gamma (1-\beta ^2)]^{ik(1+\xi)} \, . \label {fk2}
\ee
We remark that it is now possible to verify {\it a posteriori} the existence of a separator and assumption (\ref {limB}). Indeed the evaluation of the Fourier antitransform of (\ref {fk2}),
\be
\tilde Z_-(\theta)=\int _{-\infty +i\tau}^{+\infty +i\tau} \frac
{dk}{2\pi}e^{ik\theta} {\hat Z}_-(k) \, , \quad \tau >0 \, , \label {afk}
\ee
can be approximately performed, in the limit $\tilde \omega \rightarrow +\infty$, applying the residues technique after closing the contour of integration in the lower $k$-complex half-plane. The leading contributions come from the poles at $k=0$ and at $k=-2i/(1+\xi)$:
\be
\tilde Z_-(\theta)\simeq -2\beta ^2 \tilde \omega +2\pi ^{3/2}\frac {\Gamma (1-\beta ^2)}{\Gamma \left ( \frac {1}{2}+\beta ^2 \right )}\left ( \frac {\beta ^2 \tilde \omega}{2\pi} \right ) ^{2\beta ^2-1}e^{2(1-\beta ^2)\theta} \,  . \label{Zwh}
\ee
Now the condition $\tilde Z_-(\theta)=(2n+1)\pi $ identifies the holes, which happen to appear at values of $\theta$
\[
\theta _n\simeq \frac {1}{2-2\beta ^2} \ln \left [ \frac {\Gamma \left (\frac {1}{2}+\beta ^2 \right )}{2\pi ^{3/2}\Gamma (1-\beta ^2)}\left ( \frac {\beta ^2 \tilde \omega}{2\pi} \right )^{1-2\beta ^2}(2n\pi +\pi +2\beta ^2 \tilde \omega )  \right ] \, .
\]
Hence, it follows that, when $\tilde \omega \rightarrow +\infty$, all these holes go like:
\be
\theta _n \simeq \ln \tilde \omega \, .
\ee
Therefore property (\ref {limB}) holds for the separator $s(\tilde \omega)$.

\medskip

We now want to consider the vacuum eigenvalues of the transfer matrix and in particular the conserved charges defined by (\ref {In}), eventually finding the exact form (\ref {Cn1}) for the normalisation coefficients $c_{n+1}$. Formula (\ref {In}) reads
\[
c_{n+1}I_{2n+1}=\int _{-\infty}^{+\infty} d\theta \, e^{\theta (2n+1)}{\mbox {Im}}\ln \left [1+e^{i\tilde Z(\theta+i0)}\right ] \, ,
\]
and it is valid when $0\leq\tilde \omega <\pi/2\beta ^2$.
With the described procedure, we can write the expression valid also for $\tilde \omega > \pi/2\beta ^2$, i.e. when holes are present on the real axis:
\be
c_{n+1}I_{2n+1}=\frac {1}{2} \int _{-\infty}^{s(\tilde \omega)}d\theta
e^{(2n+1)\theta}\tilde Z(\theta) +
 \int _{s(\tilde \omega)}^{+\infty} d\theta \, e^{(2n+1)\theta }{\mbox {Im}}\ln \left [1+e^{i\tilde Z(\theta+i0)}\right ] \, . \label {inth}
\ee
As well as for the nonlinear integral equation, in the limit $\tilde \omega \rightarrow +\infty$ the second addendum in the r.h.s. of ({\ref {inth}) is negligible, yielding
\be
c_{n+1}I_{2n+1}\simeq \frac {1}{2}\int _{-\infty}^{s(\tilde \omega)} d\theta e^{(2n+1) \theta }
\tilde Z(\theta ) \, , \label {clin}
\ee
i.e. from (\ref {fk})
\be
c_{n+1}I_{2n+1}\simeq \frac {1}{2} {\hat Z}_-(i(2n+1)) \, .
\ee
And finally (\ref {fk2}) yields
\be
c_{n+1}I_{2n+1}\simeq -\frac {1+\xi}{(n+1)!} \frac {\beta ^{4n+4}\, \tilde \omega ^{2n+2}}{\pi ^{2n+\frac {1}{2}}\, 2^{2n+3}} \frac {\Gamma \left (\left (n+\frac {1}{2}\right )(1+\xi)\right )}{\Gamma \left (1+\left (n+\frac {1}{2}\right )\xi\right )} \, [\Gamma (1-\beta ^2)]^{-(2n+1)(1+\xi)} \, . \label{fn}
\ee
On the other hand, local conserved charges of conformal field theories can be constructed in terms of Virasoro generators \cite {FFr}. Their vacuum eigenvalues can be calculated for the first cases and we easily understand that they behave for large $\tilde \omega$ as
\be
I_{2n+1}\simeq \left ( \frac {\beta \tilde \omega}{2\pi}\right )^{2n+2} \, . \label {Ilim}
\ee
The normalisation coefficients $c_{n+1}$ of expansion (\ref {expans15}) are independent on $\tilde \omega$. Therefore, their exact value is obtained by inserting in (\ref {fn}) the expression (\ref {Ilim}) for $I_{2n+1}$. Therefore we obtain, in conclusion, the anticipated result (\ref {Cn1}).

\subsection{The case $\beta ^2 =1/2$}

When $\beta ^2 =1/2$ (free fermion point) the nonlinear integral equation (\ref {ddv}) becomes trivial, because - as follows from definition (\ref {G(x)}) - $G(x)=0$.
Its solution is
\be
Z(x)=-A\varepsilon e^{-2\varepsilon x}+\varepsilon 2\pi \omega \, .
\ee
In this case it is possible to evaluate explicitly the eigenvalue of the transfer matrix. We fix the arbitrary constant $A$ as in Remark  4 (but we do not choose $\varepsilon$) and we first consider the case in which the spectral parameter $\alpha$ is in the region 1 (\ref {reg1}):
\[
0< {\mbox {Im}} \alpha < \pi /2 \, .
\]
Some care is necessary, for the bulk terms of formul{\ae} (\ref {sumf8}) and (\ref {resumf8}) diverge logarithmically when $\beta ^2=1/2$. The analytic regularisation is again very suitable and as shown for instance in \cite {FMS} it is equivalent to a minimal renormalisation scheme: we define $\beta ^2\equiv 1/2-\eta$, and we will let $\eta \rightarrow 0$, maintaining only the finite part \footnote {More in general, divergences of the bulk terms occur for $\beta ^2=1-1/2n$, with $n\geq 1$.}. Considering first $F^+(\alpha)$, we have from (\ref {sumf8})
\be
F^+(\alpha)=B^+(\alpha)+C^+(\alpha)\, ,
\ee
where the bulk divergent part reads
\be
B^+(\alpha)=\left [\frac {\pi}{\eta} + 2\pi (1+\psi (1)-\ln \pi )+4\pi \varepsilon \alpha \right ]e^{-2\varepsilon \alpha} +o(\eta ^0) \, ,\label {Bdef}
\ee
where $\psi (1)=-{\bf C}=-0.57721566490...$ (${\bf C}$ is Euler-Mascheroni constant) is the renormalisation constant depending on the regularisation scheme (for instance, if we had considered the dependence of the function (\ref {Bdef}) on $y$, we would have obtained a different constant). Instead, the regular finite size correction is given by
\be
C^+(\alpha)=\frac {i}{\pi}\int _{0}^{+\infty}dk \frac {e^{-2\varepsilon \alpha}}{k^2-e^{-4\varepsilon \alpha}}\left [ \ln (1+e^{2\pi ^2 i e^{2i0}k-2\pi i \omega})-\ln (1+e^{-2\pi ^2 i e^{-2i0}k+2\pi i \omega})\right ] \, . \label {reCdef}
\ee
We now go to the limit $\eta \rightarrow 0$ using a minimal renormalisation scheme:
\be
B^+(\alpha)=\left [2\pi (1+\psi (1)-\ln \pi )+4\pi \varepsilon \alpha \right ]e^{-2\varepsilon \alpha}  \, . \label {reBdef}
\ee
On the contrary $C^+(\alpha)$ (\ref {reCdef}) is finite and can be explicitly calculated \cite {GR}:
\ba
C^+(\alpha)&=&-\varepsilon \ln \Gamma \left ( \frac {1}{2}+\omega +\varepsilon \pi e^{-2\varepsilon \alpha} \right ) +\varepsilon \ln \Gamma \left ( \frac {1}{2}+\omega -\varepsilon \pi e^{-2\varepsilon \alpha} \right ) -\nonumber \\
&-&\varepsilon i \pi \omega -[2\pi (1-\ln \pi)+4\pi \varepsilon \alpha -i\varepsilon \pi ^2 ]e^{-2\varepsilon \alpha} \, .
\ea
Therefore, gathering the previous two formul{\ae}, we obtain, in the region 1 (\ref {reg1}),
\ba
&&F^+(\alpha)=B^+(\alpha)+C^+(\alpha)=-\varepsilon \ln \Gamma \left ( \frac {1}{2}+\omega +\varepsilon \pi e^{-2\varepsilon \alpha}\right )+  \label {fc+} \\
&+& \varepsilon \ln \Gamma \left ( \frac {1}{2}+\omega -\varepsilon \pi e^{-2\varepsilon \alpha} \right )+2\pi \psi (1) e^{-2\varepsilon \alpha} +i\pi ^2 \varepsilon e^{-2\varepsilon \alpha} -\varepsilon i \pi \omega  \, . \nonumber
\ea
For what concerns $F^-(\alpha)$, it comes from the general relation (\ref {corr}). At $\beta ^2=1/2$, when $\alpha$ varies in region 1, $\alpha -i\pi \beta ^2$ varies exactly in region 2 ($-\pi /2 < {\mbox {Im}} \alpha < 0$). Therefore, the value of $F^-(\alpha)$ is the finite part of the $\eta \rightarrow 0$ limit of minus expression (\ref {resumf8}) in which $\alpha$ is shifted by $-i\pi \beta ^2$. Performing this limit as we did before for $F^+$, we obtain the renormalised values
\be
F^-(\alpha)=B^-(\alpha)+C^-(\alpha)+2 \varepsilon i\pi \omega \, ,
\ee
where
\be
B^-(\alpha)=B^+(\alpha)-2\pi ^2 i\varepsilon e^{-2\varepsilon \alpha} \, , \label{Bmeno}
\ee
and
\be
C^-(\alpha)=C^+(\alpha) \, . \label {Cmeno}
\ee
As a consequence,
\ba
F^-(\alpha)&=&-\varepsilon \ln \Gamma \left ( \frac {1}{2}+\omega +\varepsilon \pi e^{-2\varepsilon \alpha} \right ) +\varepsilon \ln \Gamma \left ( \frac {1}{2}+\omega -\varepsilon \pi e^{-2\varepsilon \alpha} \right ) + \nonumber \\
&+&2\pi \psi (1) e^{-2\varepsilon \alpha} -i \pi ^2 \varepsilon e^{-2\varepsilon \alpha} +\varepsilon i \pi \omega \, . \label {fc-}
\ea

Putting together the expressions for $F^+$ and $F^-$, we finally end up with   the vacuum eigenvalue of the transfer matrix when  $0< {\mbox {Im}} \alpha < \pi /2$:
\ba
{\Lambda }(\alpha)&=&{\mbox {exp}}[F^+(\alpha)] + {\mbox {exp}}[F^-(\alpha)]=\nonumber \\
&=&2\left [ \frac {\Gamma \left (\frac {1}{2}+\omega -\varepsilon \pi e^{-2 \varepsilon  \alpha} \right )}{\Gamma \left (\frac {1}{2}+\omega +\varepsilon \pi e^{-2 \varepsilon \alpha}\right)}\right ] ^{\varepsilon} e^{ 2\pi \psi (1) e^{-2\varepsilon \alpha}} \cosh (i \pi ^2 \varepsilon e^{-2\varepsilon \alpha}-\varepsilon i \pi \omega) =  \nonumber \\
&=& \frac {2\pi e^{2\pi \psi (1)e^{-2\varepsilon \alpha}}}{\Gamma \left (\frac {1}{2}+\omega +\pi e^{-2\varepsilon \alpha} \right )
\Gamma \left (\frac {1}{2}-\omega +\pi e^{-2\varepsilon \alpha} \right )} \, . \label{ffp}
\ea
This gives the expression for the vacuum eigenvalue of our conformal transfer matrix at $\beta ^2=1/2$ and in the region 1 (\ref {reg1}) of the spectral parameter $\alpha$; it is, of course, an entire function of $\lambda ^2=e^{2\alpha}$.

\medskip

In region 2 (\ref {reg2}), i.e. $-\pi /2 < {\mbox {Im}} \alpha < 0$, the calculations are carried on in a similar way. We only remark that the computation of $F^+(\alpha)$ is made starting from formula (\ref {resumf8}) and that, because of the relation (\ref {corr2})
\be
F^-(\alpha)=-F^+\left (\alpha+i\gamma \right)=-F^+\left (\alpha+i\frac {\pi}{2}+i\pi \eta \right) \, ,
\ee
the evaluation of $F^-(\alpha)$ comes from formula (\ref {sumf8}). The final result is:
\ba
F^{\pm}(\alpha)&=&-\varepsilon \ln \Gamma \left ( \frac {1}{2}+\omega +\varepsilon \pi e^{-2\varepsilon \alpha} \right ) +\varepsilon \ln \Gamma \left ( \frac {1}{2}+\omega -\varepsilon \pi e^{-2\varepsilon \alpha} \right ) + \nonumber \\
&+&2\pi \psi (1) e^{-2\varepsilon \alpha} \pm i \pi ^2 \varepsilon e^{-2\varepsilon \alpha} \mp \varepsilon i \pi \omega \, .
\ea
These expressions coincide in form with (\ref {fc+}, \ref {fc-}), therefore expression
(\ref {fc+}) for $F^+(\alpha)$ and expression (\ref {fc-}) for $F^-(\alpha)$ are valid in all the strip $-\pi/2 <{\mbox {Im}} \alpha <\pi/2$. Since $F^{\pm}(\alpha)$ are periodic with period $i\pi$ (\ref {perio}) and since (\ref {fc+}, \ref {fc-}) show the same periodicity, we conclude that in all the complex plane the expressions of $F^+(\alpha)$ and $F^-(\alpha)$ at $\beta ^2 =1/2$ are given respectively by (\ref {fc+}) and (\ref {fc-}).
It follows that the vacuum eigenvalue of the transfer matrix in all the complex plane at $\beta ^2=1/2$ is given by expression (\ref {ffp}).

In the case $\varepsilon =-1$ expression (\ref {ffp}) becomes
\ba
{\Lambda }(\alpha)=\frac {2\pi e^{2\pi \psi (1)e^{2\alpha}}}{\Gamma \left (\frac {1}{2}+\omega +\pi e^{2\alpha} \right )
\Gamma \left (\frac {1}{2}-\omega +\pi e^{2\alpha} \right )} \, ,
\label {traffp}
\ea
which means that, after the usual identifications $\lambda ^2=e^{2\alpha}$ and  (Remark 4) $\omega =2p$, ${\Lambda }(\alpha)$ coincides with the expression for the transfer matrix at $\beta ^2=1/2$ given in formula (4.32) of \cite {BLZ2}.

\medskip

Expression (\ref {ffp}) satisfies a peculiar functional equation, which identifies as its solutions all the transfer matrix eigenvalues (at the free fermion point), once we assume the solution to be entire. This functional equation is nothing but the renormalised version of the $T$-$Q$ relation which holds for $0<\beta ^2 <1/2$ \cite {BAX,BLZ2}. Indeed, we can decompose $\Lambda ^{\pm}(\alpha)$ in the following form
\be
{\mbox {exp}}[F^{\pm}(\alpha)]={\mbox {exp}}(\pm i\pi ^2 \varepsilon e^{-2\varepsilon \alpha}\mp \varepsilon i\pi \omega) \frac {Q\left (\alpha+\frac {i\pi}{2} \right )}{Q(\alpha)} \, , \label {defQ}
\ee
where we have defined
\be
Q(\alpha)\equiv \left [ \frac {\Gamma \left ( \frac {1}{2}+\omega +\pi \varepsilon  e^{-2\varepsilon \alpha}\right ) }{\Gamma \left (\frac {1}{2}+\omega \right ) } \right ]^{\varepsilon}
{\mbox {exp}}[-\pi \psi (1) e^{-2\varepsilon \alpha}] \, .
\ee
Since
\[
\frac {Q\left (\alpha-\frac {i\pi}{2}\right ) }{Q(\alpha)}= \frac {Q\left (\alpha+\frac {i\pi}{2}\right ) }{Q(\alpha)} \, ,
\]
expression (\ref {ffp}) can be written as follows
\be
Q(\alpha) {\Lambda }(\alpha)={\mbox {exp}}(i \pi ^2 \varepsilon e^{-2\varepsilon \alpha}-\varepsilon i\pi \omega) {Q\left (\alpha+\frac {i\pi}{2}\right ) } + {\mbox {exp}}(-i \pi ^2 \varepsilon e^{-2\varepsilon \alpha}+\varepsilon i\pi \omega) {Q\left (\alpha-\frac {i\pi}{2}\right ) } \, . \label{TQ}
\ee
Therefore, assuming that our model at $\varepsilon =-1$ is a lattice regularisation of the continuous theory in \cite {BLZ2}, we have proved the
functional equation (4.27) of \cite {BLZ2}. This equation has been proved there as exact at first order in perturbation theory and then used to derive (\ref {traffp}).

\medskip

Finally, we present the calculation, at the free fermion point, of the local integrals of motion. We can specialize formula (\ref {expans15}) - with $A$ chosen as in Remark 4 - or, alternatively, expand asymptotically the logarithm of (\ref {ffp}), taking into account relations (\ref {expans15}, \ref {Cn1}, \ref {In}):
\be
\ln \Lambda (\alpha)=\left [2\pi (1+\psi (1)-\ln \pi )+4\pi \varepsilon \alpha \right ]e^{-2\varepsilon \alpha}-\frac {2\varepsilon}{\pi} \sum _{n=0}^{\infty}e^{2\varepsilon \alpha(2n+1)}c_{n+1}I_{2n+1} \, ,
\ee
where
\be
c_{n+1}=-\frac {2^n}{\pi ^{2n}}\frac {1}{(n+1)(2n+1)}
\ee
and
\be
I_{2n+1}=-\varepsilon \frac {1}{2^{n+1}}B_{2n+2}\left (\frac {1}{2}+\omega \right) \, ,
\label {last}
\ee
where $B_{2m}(x)$, $m=1,2,...$, are the Bernoulli polynomials \cite {GR}. The previous formula (\ref {last}) has been already found using the free fermion basis in \cite {KK}.

\section{Summary and conclusions}

In the whole interval $0<\beta ^2 <1$, we have determined the nonlinear integral equation for the $\tilde \omega$-vacuum states of the left and right (m)KdV models, whose discretisations are dynamically twisted spin $-1/2$ XXZ chains. The mapping $\beta ^2 \rightarrow 1-\beta ^2$ connects the (m)KdV nonlinear integral equation with the nonlinear integral equation coming from the twisted spin $+1/2$ XXZ chain. We have identified conformal primary states in those vacua and the treatment of excitations -- as well as off-critical behaviour -- can be derived along the lines of \cite{FMQR}. We have given an expression for the eigenvalues of the transfer matrix in terms of solutions of the nonlinear integral equation in the whole interval $0<\beta ^2 <1$, and we have found a rich analytic structure, which in particular implies different asymptotic expansions. From the latter the expressions of the local integrals of motion shared with KdV theory have been extracted. At the free fermion point the aforementioned eigenvalues are evaluated exactly, again in agreement with \cite {BLZ2}. Therefore, we have been naturally led to conclude that our model furnishes an effective framework for calculating interesting quantities of both quantum KdV theory and Conformal Field Theories. Moreover, the exact values of the transfer matrix eigenvalues may be useful in ${\cal P}{\cal T}$- symmetric Quantum Mechanics applications, for they have been proved to be spectral determinants in \cite {DT}. In this respect, the exact solution of the theory given in the limit of infinite twist ($\tilde \omega \rightarrow +\infty$) may be of particular interest in many applications and developments. Finally, we would like to highlight that the second paper will concern the off-critical theory approach proposed in \cite{FR} within this framework, showing the generality of the methodology.

\medskip

{\bf Acknowledgements} - We thank P. Dorey and R. Tateo for comments
and interest in this work. D.F. thanks EPRSC (grant GR/M66370) and
Leverhulme Trust (grant F/00224/G). M.R. thanks EPRSC for the grant
GR/M97497 and  the Department of Mathematical Sciences of Durham for
warm hospitality. This work has been also supported by EC FP5
Network, contract number HPRN-CT-2002-00325.

\end{document}